%% file: ex_article.tex
\documentclass[final,hidelinks,onefignum,onetabnum]{siamart220329}

\input{ex_shared}
\input{new_commands}

\ifpdf
\hypersetup{
  pdftitle={},
  pdfauthor={}
}
\fi

\begin{document}
					
\maketitle

\begin{abstract}\input{sections/abstract.tex}
\end{abstract}

\begin{keywords}
preconditioning, hyper-power series, iterative methods, matrix-free, saddle point problem, Stokes equations, isogeometric analysis
\end{keywords}

\begin{MSCcodes}
65N30, 65F08, 65F10, 65F30
\end{MSCcodes}

\input{sections/section1_introduction}

\input{sections/section2_method.tex}

\input{sections/section2_model.tex}

\input{sections/section3_preconditioner.tex}

\input{sections/section4_benchmarks.tex}

\input{sections/section5_conclusion.tex}


\bibliographystyle{siamplain}
\bibliography{references}

\appendix

\input{sections/appendix1_diagonalization.tex}

\input{sections/appendix2_discretization.tex}

\end{document}

%% file: ex_shared.tex
\usepackage{lipsum}
\usepackage{amsfonts}
\usepackage{amssymb}
\usepackage{graphicx}
\usepackage{caption}
\DeclareCaptionFormat{cont}{#1 (cont.)#2#3\par}
\usepackage{epstopdf}
\usepackage{cases}
\usepackage{tikz}
\usepackage{tikz-cd}
\usepackage{diagbox}
\usepackage{multirow}
\usepackage{algorithmicx}
\usepackage{hyperref}
\usepackage{tikz}
\usepackage{multirow}
\usepackage{nicematrix}
\usepackage{algpseudocode}
\usepackage{lscape} 
\Crefname{ALC@unique}{Line}{Lines}
\ifpdf
  \DeclareGraphicsExtensions{.eps,.pdf,.png,.jpg}
\else
  \DeclareGraphicsExtensions{.eps}
\fi

\usetikzlibrary{calc}
\usetikzlibrary{positioning}
\usetikzlibrary{decorations.markings,shadings}
\usetikzlibrary{calc,intersections,patterns,decorations.pathmorphing,patterns.meta,arrows.meta}
\tikzset{
  myarrow/.style={
    ->,
    shorten >=1pt,
    shorten <=1pt,
    line width=0.5pt,
    arrow fill=black!50,
    arrow fill opacity=0.5,
    arrow head=0.1cm,
    arrow body=0.2cm,
  }
}

\tikzdeclarepattern{
  name=arrows,
  type=uncolored,
  bottom left={(-.1pt,-.1pt)},
  top right={(12.1pt,8.1pt)},
  tile size={(12pt,8pt)},
  tile transformation={rotate=0},
  code={
\tikzset{x=1pt,y=1pt}
\draw [-{Latex[length=2pt, width=2pt]}, line width=0.1pt] (0,2) -- (6,2); 
\draw [-{Latex[length=2pt, width=2pt]}, line width=0.1pt] (6,6) -- (12,6); 
} }

\newsiamremark{remark}{Remark}
\newsiamremark{hypothesis}{Hypothesis}
\crefname{hypothesis}{Hypothesis}{Hypotheses}
\newsiamthm{claim}{Claim}

\headers{Preconditioning using the hyper-power method}{M.\L. Mika, M.F.P. ten Eikelder, D. Schillinger and R.R. Hiemstra}

\title{Matrix-free polynomial preconditioning of saddle point systems using the hyper-power method}

\author{M.\L{}. Mika\thanks{Corresponding author (\email{michal.mika@tu-darmstadt.de})}
\and M.F.P. ten Eikelder
\and D. Schillinger
\and R.R. Hiemstra}

\usepackage{amsopn}


%% file: new_commands.tex
\newcommand{\List}[1]{\left(\right. #1 \left.\right) }					
\newcommand{\Set}[1]{\left\{\right. #1 \left.\right\} }					

\newcommand{\BigO}[1]{\mathcal O({#1})} 

\newcommand{\vectorfield}[1]{\boldsymbol{#1}}
\newcommand{\vect}[1]{\mathbf{#1}} 
\newcommand{\mat}[1]{\mathbf{#1}}
\newcommand{\mmat}[1]{\mathsf{#1}}

\newcommand{\domain}{\Delta}													
\newcommand{\n}{n}																	

\newcommand{\A}{\mat A}																	
\newcommand{\B}{\mat B}																	
\newcommand{\M}{\mathcal M}																	
\newcommand{\Pc}{\mathcal P}																	

\newcommand{\p}{p}																	
\renewcommand{\r}{r}																	

\newcommand{\sspace}[2]{\mathbb{S}^{#1}_{#2}}					
\newcommand{\Span}[1]{\text{span} \left( #1 \right)}					

\newcommand{\oplushat}{\mathop{\hat{\oplus}}}		

\newcommand{\spacedimension}[1]{\mathrm{dim}\,#1}

\newcommand{\matzero}{\cdot}

\newcommand{\PinvA}{\mat P}

%% file: sections/abstract.tex
This study explores the integration of the hyper-power sequence, a method commonly employed for approximating the Moore-Penrose inverse, to enhance the effectiveness of an existing preconditioner. The approach is closely related to polynomial preconditioning based on Neumann series. We commence with a state-of-the-art matrix-free preconditioner designed for the saddle point system derived from isogeometric structure-preserving discretization of the Stokes equations.  Our results demonstrate that incorporating multiple iterations of the hyper-power method enhances the effectiveness of the preconditioner, leading to a substantial reduction in both iteration counts and overall solution time for simulating Stokes flow within a 3D lid-driven cavity. Through a comprehensive analysis, we assess the stability, accuracy, and numerical cost associated with the proposed scheme.

%% file: sections/section1_introduction.tex
\section{Introduction}\label{sec:introduction}
Preconditioning is an essential aspect of iterative solution methods for linear systems of equations $\mat A \vect x = \vect b$. We refer to the review papers in \cite{benzi2002preconditioning, wathen2015preconditioning} and the monographs \cite{chen2005matrix, saad2003iterative, vorst2003iterative} for an in-depth treatment. Preconditioner design is typically a trade-off between its \emph{effectiveness} to reduce the number of iterations required to reach a set tolerance and its \emph{efficiency} measured by its formation and application cost relative to the \emph{forward problem} (one application of $\mat{A}$ to a vector). In addition, a good preconditioner is robust, which means it remains effective over a wide range of model, discretization and material parameters \cite{benzi2002preconditioning, wathen2015preconditioning}.

Many Krylov methods \cite{vorst2003iterative} have a cost that scales proportionally to the number of iterations. Important examples are the \emph{Conjugate Gradient} (CG) method and the \emph{Minimal Residual} (MINRES) method. The \emph{Generalized Minimal Residual} method (GMRES) is a notable exception as its cost per iteration increases with every iteration~\cite{vorst2003iterative}. In the case of CG and MINRES the solution time can be estimated as
\begin{equation}
    \label{eq:solution_time_model}
    T_{\rm  sol} \propto N_{\rm iter} \, T_f \, ( 1 + T_p / T_f ),
\end{equation}
where $N_{\rm iter}$ denotes the number of iterations to achieve a set tolerance, $T_f$ is the time it takes to evaluate the forward problem, and  $T_p$ denotes the time it takes to apply the preconditioner to a vector. 
\begin{figure}
    \centering
    \includegraphics[trim={0cm 0cm 0cm 0cm},clip,width=0.85\textwidth]{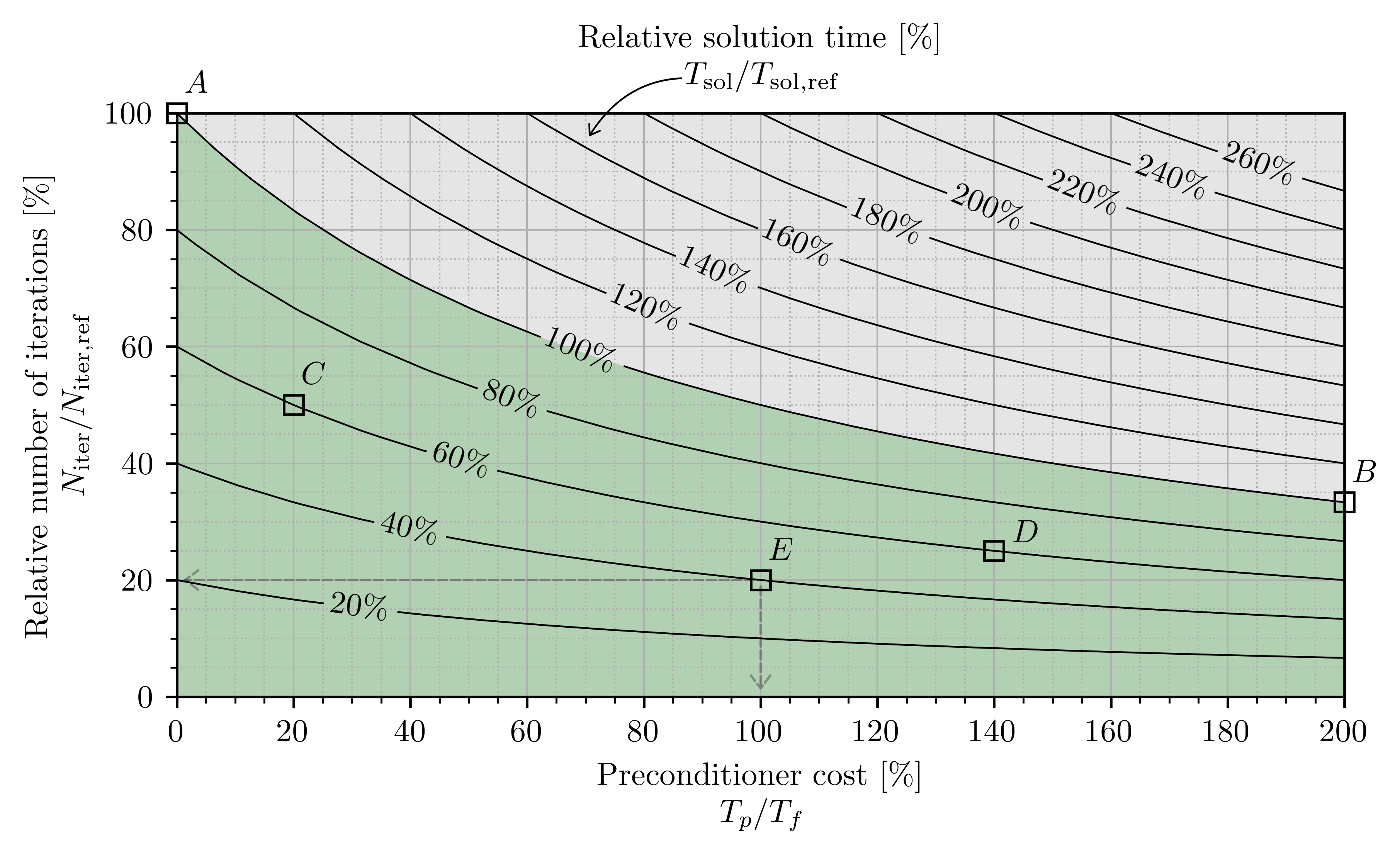}
    \vspace{-0.5cm}
    \caption{Total solution time as a function of the preconditioner cost $T_p / T_f$ (efficiency) and the number of iterations (effectiveness) compared to a reference method. The green area marks the region, in which the preconditioned system performs better than the reference method.}
    \label{fig:Pc_efficiency}
\end{figure}
Figure \ref{fig:Pc_efficiency} illustrates how solution time is dependent on preconditioner efficiency ($T_p / T_f$) and its effectiveness ($N_{\rm iter} / N_{\rm iter,ref}$) based on the relationship in Equation \eqref{eq:solution_time_model}. Here, the reference method refers to an iterative method of choice, unconditioned or conditioned, that yields a unique solution up to the specified tolerance. The green area delineates the region in which an iterative solver, when applied to the preconditioned system, outperforms or matches the performance of the reference system. The trade-off between efficiency and effectiveness can be summarized as follows: When the cost of a preconditioner is twice that of solving the forward problem, it can yield improved performance, provided that its effectiveness reduces the number of iterations by more than 67\% (point B). Conversely, a preconditioner corresponding to point C, which incurs only 20\% of the forward problem's cost, reduces the number of iterations by 50\% and matches the solution time performance of a preconditioner at point D, which is seven times more expensive but reduces the number of iterations by 75\%.

In this work we design a sequence of matrix-free preconditioners based on a self-correcting method of successive approximation. The hyper-power method of order two (Schulz \cite{Schulz1933}) is an iterative technique to approximate an ordinary inverse matrix $\mat{A}^{-1}$ via the recursion
\begin{equation}
\begin{aligned}
	\mat{X}_0 = \mat{X}, \qquad
	\mat{X}_{k+1} = 2\mat{X}_{k} - \mat{X}_{k} \mat{A} \mat{X}_{k}, \quad k = 0, \; 1, \, \ldots \,.
	\label{eq:hyperpower2}
\end{aligned}
\end{equation}
The sequence converges uniformly with order two to $\mat{A}^{-1}$ if the initial matrix $\mat{X}$ satisfies the condition $\rho(\mat{I} - \mat{A} \mat{X}) < 1$, where $\rho$ denotes the spectral radius. This follows directly if one considers the equivalent Neumann-like sequence \cite{ClThWe2001, Tanabe1975} 
\begin{equation}
\begin{aligned}
	\mat{X}_{k+1} = \sum_{j=0}^{2^{k+1}-1} \left( \mat{I} -  \mat{X}  \mat{A}  \right)^j \mat{X}, \quad k = 0, \; 1, \, \ldots \, .
	\label{eq:neumann}
\end{aligned}
\end{equation}
The method was extended and studied in the context of generalized inverses by Ben-Israel \cite{BeIsAd1965} and others. Hyper-power methods of higher order are well described in \cite[Chapter 7.7, Page 270]{ben2003generalized} and \cite[Chapter 4.1, page 94]{householder2013theory} and the references cited therein. 

It follows from \eqref{eq:neumann} that $k$ steps of our approach are equivalent to a truncated Neumann series of order $2^k$. Polynomial preconditioning, e.g. based on Neumann series, is a well known technique to accelerate convergence of Krylov methods \cite{vorst2003iterative} at the expense of multiple matrix-vector products per iteration \cite{DuGrRo1979,JoMiChaGe1983,OLeary1991,Saad1985,Saad1989}. The reduction in the number of iterations cannot exceed a factor more than the order of the matrix polynomial, which means that the total number of matrix-vector products cannot be reduced \cite{OLeary1991}. However, the remaining operations in a Krylov method are reduced (such as vector inner products), which can still lead to substantial savings, particularly on high performance distributed systems \cite{Saad1985,Saad1989}. Instead of using $\mat{A}$ in \eqref{eq:hyperpower2}, we use an accurate approximation that enables more efficient matrix vector multiplications. Hence, although the total number of matrix-vector products cannot be reduced, they are far less expensive, leading to an overall reduction in solution time. Despite the fact that evaluating the equivalent truncated Neumann series is equal in terms of the number of matrix-vector products, we present our work from the perspective of the iterative second-order hyper-power method. We remark that other more powerful iterative schemes for the computation of outer-inverses exist in the literature \cite{petkovic2015hyper}.

The strategy is applied to a saddle point system arising from structure preserving isogeometric discretization of the Stokes equations \cite{BuFaSa2010, EvHu2013, HiToHuGe2014}. Krylov methods tend to converge rather slowly when applied to saddle point systems \cite{benzi2008saddlepoint}. The objective is to attain a preconditioner that achieves a better trade-off between effectiveness and efficiency. To achieve this we do the following. The sequence is initialized with a state-of-the-art matrix-free block-diagonal preconditioner in \cite{MoSaTa2018}. In each step of the recurrent scheme we generate a matrix-free block-diagonal preconditioner based on an accurate approximation of the matrix (rather than the matrix $\mat{A}$ in \eqref{eq:hyperpower2} itself) and the previous preconditioner in the sequence. The approximation exploits Kronecker structure \cite{Loan2000}, which enables highly efficient matrix-vector products, leading to a better trade-off between preconditioner effectiveness and application cost. 

The outline of this paper is as follows. In Section \ref{sec:method}, we describe our method to improve the trade-off between preconditioner effectiveness and cost as well as its properties. In Section \ref{sec:model}, we introduce the saddle point system arising from isogeometric structure preserving discretization of the Stokes equation. The design of preconditioners using the presented approach and a discussion of their application cost is provided in Section \ref{sec:preconditioning_strategy}. The application to the lid-driven cavity benchmark is presented in Section \ref{sec:benchmark}. The contributions and results are summarized in Section \ref{sec:conclusion} together with recommendations for future work.

%% file: sections/section2_method.tex
\section{Preconditioners based on the hyper-power method}
\label{sec:method}
We construct a sequence of preconditioners $\Pc_{k} \in \mathbb{R}^{N\times N}$, $k \geq 0$, for $\mat{A}$ based on the recurrence relation in Equation \eqref{eq:hyperpower2}. More precisely, suppose that we have an initial positive definite preconditioner $\Pc_0$ and an approximation $\tilde{\mat A}$ of $\mat{A}$, such that $\kappa(\tilde{\mat{A}}^{-1}\mat{A}) < \kappa(\Pc_0^{-1}\mathbf{A})$, where $\kappa(\cdot)$ denotes the condition number of a matrix. The sequence of preconditioners is given by
\begin{align}\label{eq:hyperpowersequence}
    \Pc_{k+1}^{-1} = 2 \Pc_{k}^{-1} - \Pc_{k}^{-1} \tilde{\mat{A}} \Pc_{k}^{-1}.
\end{align}
Note that at no point the inverse of $\tilde{\mat A}$ is used in the scheme. This enables considerable flexibility in the choice of the approximation.

Let us discuss the properties of the sequence in Equation \eqref{eq:hyperpowersequence}. We define $\PinvA_k = \Pc^{-1}_k \tilde{\mat{A}}$, and assume that the spectrum $\sigma\left(\PinvA_{0}\right) \subset (0,2)$.
\begin{lemma}
  \label{lemma:Sk_posdef}
  The matrices $\PinvA_k$ are positive definite for $k \geq 0$ and in particular the spectrum $\sigma\left(\PinvA_{k}\right) \subset (0,1]$ for $k > 0$.
\end{lemma}
\begin{proof}
    The positive eigenvalues imply the positive definiteness of $\PinvA_0$. We denote its eigendecomposition as $\PinvA_0 = \mat U_0 \mat\Lambda_0 \mat U_0^{-1}$, where $\mat U_0$ is a unitary matrix of which the columns form an orthonormal basis of the eigenvectors of $\PinvA_0$ and~$\mat\Lambda_0$ is a diagonal matrix with the eigenvalues of $\PinvA_0$ on the diagonal. A~straightforward substitution of the sequence \eqref{eq:hyperpowersequence} reveals
    \begin{align}\label{eq:S1form}
      \PinvA_1 = \mat U_0 \left( 2\mat\Lambda_0 - \mat\Lambda_0^2 \right)\mat U_0^{-1}.
    \end{align}
    As a consequence of \eqref{eq:S1form}, we obtain $\sigma\left(\PinvA_1\right) \subset (0,1]$, where we have invoked the assumption $\sigma\left(\PinvA_0\right) \subset (0,2)$. Using the same argument, we deduce
    \begin{align}\label{eq:Skform}
      \PinvA_{k+1} = \mat U_k \left( 2\mat\Lambda_k - \mat\Lambda_k^2 \right)\mat U_k^{-1},
    \end{align}
    for $k \geq 0$. Thus, $\sigma\left(\PinvA_k\right) \subset (0,1]$ and hence $\PinvA_k$ are positive definite for $k \geq 0$. 
\end{proof}

We now study the extreme values of the spectrum. To this purpose, let us denote $\lambda_{k,{\rm min}}= \min\, \sigma \left(\PinvA_k\right)$ and $\lambda_{k,{\rm max}}= \max\, \sigma \left(\PinvA_k\right)$. Additionally, motivated by \eqref{eq:Skform}, we define $l\;:\;(0,2) \rightarrow  (0,1]$ as $l(\lambda) = 2\lambda - \lambda^2$.
\begin{corollary}
  \label{cor:Skform}
  We have $\sigma(\PinvA_{k+1})=\Set{l(\lambda) ~|~ \lambda \in \sigma(\PinvA_k)}$. Additionally, for the minima and maxima of $\sigma(\PinvA_{k+1})$, we have
  \begin{itemize}
    \item $\lambda_{1,{\rm min}} = {\rm min}\Set{ l(\lambda)~|~\lambda \in \Set{ \lambda_{0,{\rm min}},\lambda_{0,{\rm max}} } } $.
    \item $\lambda_{1,{\rm max}} = {\rm max}\Set{ l(\lambda)~|~\lambda \in \Set{ \lambda_{0,{\rm min}},\lambda_{0,{\rm max}} } } $.
    \item $\lambda_{k+1,{\rm max}} \leq 1$ for $k \geq 0$. 
    \item $\lambda_{k+1,{\rm min}} = l(\lambda_{k,{\rm min}})$ for $k \geq 1$.
  \end{itemize} 
\end{corollary}
\begin{proof}
  The form of $\sigma(\PinvA_{k+1})$, $k \geq 0$, follows from the identity \eqref{eq:Skform}. The other results are a consequence of $l\;:\;(0,2) \rightarrow  (0,1]$ being a concave function. 
\end{proof} 
\begin{remark}
  By virtue of Corollary \ref{cor:Skform} we can a priori predict the effectiveness after each update. We merely need to compute the extreme eigenvalues of $\PinvA_0$.
\end{remark}

Next, we introduce a sequence of matrices $\tilde{\PinvA}_k := \Pc_k^{\frac 1 2}\PinvA_{k}\Pc_k^{-\frac 1 2}$, $k \geq 0$. 
\begin{proposition}
  \label{prop:samespectrum}
  The matrices $\PinvA_k$ and $\tilde{\PinvA}_k$ have the same spectrum.
\end{proposition}
\begin{proof}
  Given that $\Pc_k$ is invertible, $\PinvA_k$ and $\tilde{\PinvA}_k$ are similar and thus share the same spectrum.
\end{proof}

We are now ready to prove the positive definiteness of each $\Pc_{k}$.
\begin{theorem}\label{thm:positive definite}
  The preconditioners $\Pc_k$, $k \geq 0$, are positive definite.
\end{theorem}
\begin{proof}
  The matrix $\Pc_0$ is positive definite. The matrices $\Pc_{k+1}^{-1}$ may be written as
  \begin{align}
    \Pc_{k+1}^{-1} = \Pc_{k}^{-\frac 1 2} \left( 2 \mat I - \tilde{\PinvA}_k \right) \Pc_{k}^{-\frac 1 2}, \quad k \geq 0.
  \end{align}
  Hence, $\Pc_{k+1}^{-1}$ is congruent to $2 \mat{I} - \tilde{\PinvA}_k$. From Lemma \ref{lemma:Sk_posdef} and Proposition \ref{prop:samespectrum} we find
  \begin{align}
      \sigma\left(2 \mat{I} - \tilde{\PinvA}_k\right) \subset (0,2)\quad\rm for \quad k \geq 0.
  \end{align}
  Invoking Sylvester's law of inertia provides that the spectra of both matrices share the same signature, i.e. the number of positive, negative and zero eigenvalues is equal. Hence, all eigenvalues of $\Pc_{k+1}^{-1}$ are positive and $\Pc_k$ is positive definite for $k\geq 0$. 
\end{proof}
\begin{remark}\label{rmk:convergence_Pk}
  The lower bound on $\sigma(\PinvA_0)$ corresponds to the condition necessary for the convergence of \eqref{eq:hyperpowersequence}, which is $\max\,\sigma(\mat I - \PinvA_k) < 1$ \cite{PeSt2011}. The upper bound ensures that each element in the sequence $\mathcal P_k$ is positive definite. In practice, this translates to choosing a \emph{good enough} initial preconditioner for $\tilde{\mat A}$ in order to ensure positive definiteness.
\end{remark}

To study the effectiveness of the updates, we denote the condition number of $\PinvA_{k}$ in the standard way as $\kappa(\PinvA_{k}):=\lambda_{k,{\rm max}}/\lambda_{k,{\rm min}}$.
\begin{corollary}
  \label{cor:decreasing_cond_numbers}
  The condition numbers $\kappa(\PinvA_{k}) > 1$, $k > 0$, form a strictly decreasing sequence.
\end{corollary}
\begin{proof}
 Let us introduce the slope $s\;:\;(0,1] \rightarrow \mathbb{R}$ of $l$ relative to the origin,
  \begin{align}\label{eq:s_form}
    s(\lambda) := \frac{l(\lambda)}{\lambda} 
  \end{align}
  Since $l$ is concave and strictly increasing, we deduce that $s$ is monotonically decreasing in $\lambda$. Given that $\kappa(\PinvA_k) > 1$, we conclude
  \begin{align}
    s(\lambda_{k,{\rm max}}) < s(\lambda_{k,{\rm min}}).
  \end{align}
  Invoking Corollary \ref{cor:Skform} and inserting the form \eqref{eq:s_form} implies 
  \begin{align}
    \frac{\lambda_{k+1,{\rm max}}}{\lambda_{k,{\rm max}}} < \frac{\lambda_{k+1,{\rm min}}}{\lambda_{k,{\rm min}}}.
  \end{align}
  Recalling that $\PinvA_k$ has positive eigenvalues, we conclude
  \begin{align}
    \kappa(\PinvA_{k+1}) < \kappa(\PinvA_k)\quad \rm for \quad k > 0.
  \end{align}
\end{proof}

%% file: sections/section2_model.tex
\section{The model problem and discretization} 
\label{sec:model}
In this section we introduce the weak form of the Stokes problem that leads to the saddle point system in Equation~\eqref{eq:systemmatrix}. The weak form is discretized using structure preserving isogeometric Raviart-Thomas spaces \cite{BuFaSa2010,EvHu2013,HiToHuGe2014}. The Kronecker properties of each of the matrices are shortly discussed. We refer to Appendix \ref{app:kronecker} for a brief overview of the Kronecker product and its properties and to Appendix \ref{app:discretization} for a discussion about structure preserving isogeometric spaces.

\subsection{The model problem}

Let $\Omega \subset \mathbb{R}^3$ be an open set with a piecewise smooth boundary $\partial \Omega$ and an outward unit normal vector $\vectorfield{n}$. The domain $\Omega$ is filled with an incompressible fluid with unit density and constant viscosity $\nu$. A mixed variational formulation of the Stokes problem is the following one \cite{arnold2018finite, EvHu2013}. Let $L^2(\Omega)$ denote the set of square integrable functions on $\Omega$ and $H^1(\Omega)$ the set of functions in $L^2(\Omega)$ with derivatives in $L^2(\Omega)$. In addition, let $\vect{L}^2(\Omega)$ and $\vect{H}^1(\Omega)$ denote their vectorial counterparts. We consider vector fields and functions in the following spaces,
\begin{equation}
    \begin{aligned}
        V &:= \left\{ \vectorfield{v} \in \vect{H}^{1}(\Omega) \, \big\vert\,  \vectorfield{v} \cdot \vectorfield{n} = 0 \text{ on } \partial \Omega    \right\}, \\
        Q &:= \left\{ \right. q \in L^2(\Omega) \,\big\vert\, \int_\Omega q\,\mathrm d\Omega = 0  \left. \right\}.
        \label{eq:spaces}
    \end{aligned}
\end{equation}
The strongly imposed boundary condition on the velocity enforces no penetration through the boundary of the domain. The constraint on the pressure space is needed for well-posedness. The variational formulation reads:\\
\textit{Find $\vectorfield{u} \in V$ and $p \in Q $ such that}
\begin{equation}
    \begin{aligned}
    a(\vectorfield{u}, \vectorfield{v}) + \sigma(\vectorfield{u}, \vectorfield{v}) - b(\vectorfield{v}, p) 	&= l(\vectorfield{v})  && \forall \vectorfield{v} \in V,								\\
    b(\vectorfield{u}, q) &= 0 && \forall q \in Q,
    \end{aligned}
\end{equation}
where $a(\cdot , \cdot)$, $b(\cdot , \cdot)$, $\sigma(\cdot , \cdot)$ and $l(\cdot)$ are defined by
\begin{equation*}
    \begin{aligned}
    		a(\vectorfield{w}, \vectorfield{v}) &:= \int_{\Omega} 2\nu \, \nabla^s\vectorfield{w}  \, : \, \nabla^s \vectorfield{v} \, d \Omega, \\
    		b(\vectorfield{w}, q) 	&:= \int_{\Omega} \nabla \cdot \vectorfield{w} \, q \, d \Omega, \\
    		l(\vectorfield{v})			&:= \int_{\Omega} \vectorfield{b} \cdot \vectorfield{v} \, d \Omega, \\
    		\sigma(\vectorfield{w}, \vectorfield{v}) &:= \int_{\partial\Omega}
        2\nu \left(\alpha \vectorfield{w}\cdot \vectorfield{v}
        - ((\nabla^s \vectorfield{w})\vectorfield{n})\cdot \vectorfield{v}
        - ((\nabla^s \vectorfield{v})\vectorfield{n})\cdot \vectorfield{w}\right) \,\mathrm d \vect s. \\
    \end{aligned}
\end{equation*}
Note, that while the zero normal component of the velocity at the boundary -- the no-penetration boundary condition -- is prescribed strongly, the zero tangential component -- the no-slip boundary condition -- is imposed weakly using Nitsche's method with contributions in $\sigma(\cdot,\cdot)$ ($\alpha > 0$). These are, from left to right, the penalty term, the consistency term and the symmetry term \cite{EvHu2013}.

\subsection{Discretization}
Let $V_h \subset V$ and $Q_h \subset Q$ denote finite dimensional subspaces of \eqref{eq:spaces}, discretized using isogeometric Raviart-Thomas spaces, see Appendix~\ref{app:discretization}. Furthermore, let $\alpha = C_{pen}/h$, where $C_{\rm pen}$ is the penalty term and $h$ the mesh size. The Galerkin method yields the finite dimensional variational formulation: \\
\textit{Find $\vectorfield{u}_h \in V_h$ and $p_h \in Q_h $ such that}
\begin{equation}
    \begin{aligned}
    a(\vectorfield{u}_h, \vectorfield{v}_h) + \sigma(\vectorfield{u}_h, \vectorfield{v}_h) - b(\vectorfield{v}_h, p_h) 	&= l(\vectorfield{v}_h)  && \forall \vectorfield{v}_h \in V_h,								\\
    b(\vectorfield{u}_h, q_h) &= 0 && \forall q_h \in Q_h.
    \end{aligned}
\end{equation}

The discretization leads to the saddle point system in Equation \eqref{eq:systemmatrix}. The blocks of the saddle point matrix $\M$ are $\A\in \mathbb{R}^{n_V \times n_V}$ and $\B \in \mathbb{R}^{n_V \times n_Q}$ with the dimensions $n_V := \spacedimension{V_{h}}$ and $n_Q := \spacedimension{Q_{h}}$.
\begin{equation}
    \begin{bmatrix}
        \A & \B \\
        \B^T & \mat 0 \\
    \end{bmatrix}
    \begin{bmatrix}
        \mathbf u \\
        \mathbf p \\
    \end{bmatrix} =
    \begin{bmatrix}
        \mathbf f \\ 
        \mat 0
    \end{bmatrix}.
    \label{eq:systemmatrix}
\end{equation}
Matrix $\A$ is symmetric positive definite and matrix $\B$ is a tall matrix of full rank. Consequently, the Schur complement of $\M$ with respect to the block $\A$ exists and is given by $\mat S = -\B^T \A^{-1} \B$. We note that the pressure constraint in $Q_{h}$ does not need to be implemented if an iterative solver on Krylov subspaces is used.

Due to the tensor product nature of the approximation spaces the matrices $\A$ and $\B$ have important Kronecker structure on structured Cartesian grids and a constant uniform viscosity coefficient. Even on curvilinear meshes or non-uniform viscosity the operators can be well approximated using Kronecker product techniques \cite{MoSaTa2018}. We refer to Appendix \ref{app:kronecker} for a brief overview of the Kronecker product and Kronecker sum, which are used in the next section to construct the preconditioner from \cite{MoSaTa2018}.

\subsection{Kronecker structure on Cartesian grids}
Assuming constant parameters and Cartesian grids, the blocks in $\M$ can be expressed as (block-wise) sums of Kronecker products of univariate positive definite mass matrices $\mmat M_k$ and $\mmat{\check M}_k$, univariate positive semi-definite stiffness matrices $\mmat K_k$ and $\mmat{\check K}_k$, univariate skew-symmetric matrices $\mmat C_k$ and $\mmat{\check C}_k$, as well as matrices arising from evaluation of the Nitsche terms, $\mmat{\check{N}}_k$ and $\mmat{\check{B}}_k$. The subscript $k$ denotes parametric direction and the check accent is used to distinguish matrices of similar structure but using basis functions of different degree. The definitions of these univariate matrices are given in Appendix \ref{subsec:univariatemats}.

With the definition of a positive definite univariate matrix $\mmat{\check T}_k$,
\begin{equation*}
\mmat{\check{T}}_k = \frac 1 2 \left( \mmat{\check{K}}_k + \frac{2 C_{pen}}{h} \mmat{\check{N}}_k - \mmat{\check{B}}_k - \mmat{\check{B}}_k^T \right),
\end{equation*}
the diagonal blocks of the matrix $\mat A$ can be written in a compact form as
\begin{align*}
	\mat{A}_{11} &= 
		\mmat{\check{M}}_3 \otimes \mmat{\check{M}}_2 \otimes \mmat{K}_1 +  
		\mmat{\check{M}}_3 \otimes \mmat{\check{T}}_2 \otimes \mmat{M}_1 + 
		\mmat{\check{T}}_3 \otimes \mmat{\check{M}}_2 \otimes \mmat{M}_1 = 
		\mmat{\check T}_{3} \oplushat \mmat{\check T}_{2} \oplushat \mmat K_{1}, \\
	\mat{A}_{22} &= 
		\mmat{\check{M}}_3 \otimes \mmat{M}_2 \otimes \mmat{\check{T}}_1 +  
		\mmat{\check{M}}_3 \otimes \mmat{K}_2 \otimes \mmat{\check{M}}_1 + 
		\mmat{\check{T}}_3 \otimes \mmat{M}_2 \otimes \mmat{\check{M}}_1 = 
		\mmat{\check T}_{3} \oplushat \mmat K_{2} \oplushat \mmat{\check T}_{1}, \\
	\mat{A}_{33} &= 
		\mmat{M}_3 \otimes \mmat{\check{M}}_2 \otimes \mmat{\check{T}}_1 +  
		\mmat{M}_3 \otimes \mmat{\check{T}}_2 \otimes \mmat{\check{M}}_1 + 
		\mmat{K}_3 \otimes \mmat{\check{M}}_2 \otimes \mmat{\check{M}}_1 = 
		\mmat K_{3} \oplushat \mmat{\check T}_{2} \oplushat \mmat{\check T}_{1},
\end{align*}
and the off-diagonal blocks as
\begin{align*}
	&\mat{A}_{12} = \mmat{\check{M}}_3 \otimes \mmat{C}^T_2 \otimes \mmat{C}_1,&
	&\mat{A}_{13} = \mmat{C}^T_3 \otimes \mmat{\check{M}}_2 \otimes \mmat{C}_1,&
	&\mat{A}_{23} = \mmat{C}^T_3 \otimes \mmat{C}_2 \otimes \mmat{\check{M}}_1.&
\end{align*}
The blocks of the operator $\mat B$ can be expressed as
\begin{align*}
	&\mat{B}_{11} = \mmat{\check{M}}_3 \otimes \mmat{\check{M}}_2 \otimes \mmat{\check C}_1, &
	&\mat{B}_{12} =\mmat{\check{M}}_3 \otimes \mmat{\check C}_2 \otimes \mmat{\check{M}}_1, &
	&\mat{B}_{13} = \mmat{\check C}_3 \otimes \mmat{\check{M}}_2 \otimes \mmat{\check{M}}_1.&
\end{align*}
The Kronecker product structure presented above is essential in the design of our preconditioners as it reduces the computational cost of a matrix-vector product from $\BigO{N^2}$ for a general dense matrix to $\BigO{N^{4/3}}$ for a Kronecker product matrix of size $N \times N$.

%% file: sections/section3_preconditioner.tex
\section{Preconditioning} 
\label{sec:preconditioning_strategy}

For the system in Equation \eqref{eq:systemmatrix} we propose a sequence of matrix-free block-diagonal preconditioners $\List{\Pc_0, \Pc_1, \Pc_2,\, \ldots}$,
\begin{equation}
    \Pc_k = \begin{bmatrix}
        \Pc_{V,k} & \cdot \\
        \cdot   & \Pc_{Q,k}
    \end{bmatrix},
    \label{eq:blockdiagprecond}
\end{equation}
where $\Pc_{V,k}$ is a preconditioner for the block $\A$ and $\Pc_{Q,k}$ is a preconditioner for the (negative) Schur complement $\B^T \A^{-1} \B$. We initialize the sequence using the Kronecker product preconditioners proposed in \cite{MoSaTa2018}. The preconditioner $\Pc_{V,0}$ is based on fast diagonalization of the diagonal blocks of $\A$,
\begin{equation}
    \Pc_{V,0} = \begin{bmatrix}
        \mmat{\check T}_{3} \oplushat \mmat{\check T}_{2} \oplushat \mmat K_{1} & \matzero                                                                & \matzero \\
        \matzero                                                                      & \mmat{\check T}_{3} \oplushat \mmat K_{2} \oplushat \mmat{\check T}_{1} & \matzero \\
        \matzero                                                                      & \matzero                                                                & \mmat K_{3} \oplushat \mmat{\check T}_{2} \oplushat \mmat{\check T}_{1}
    \end{bmatrix}.
\end{equation}
The preconditioner $\Pc_{Q,0}$ is based on the Kronecker product pressure mass-matrix, 
\begin{equation}
    \Pc_{Q,0} = \frac 1 \nu \mmat{\check{M}}_3 \otimes \mmat{\check{M}}_2 \otimes \mmat{\check{M}}_1.
\end{equation}
In the following, we present how the proposed sequence is generated based on the ideas presented in Section \ref{sec:method}.

\subsection{Application to block-diagonal preconditioners}
We apply the hyper-power method in Equation \eqref{eq:hyperpowersequence} to the block diagonal preconditioner $\Pc_0$ in a block-wise manner. We start with the block $\Pc_{V,0}$. Since we work on Cartesian grids with constant parameters and matrix-vector products with $\A$ and $\Pc^{-1}_{V,0}$ have the same cost order, we approximate $\A^{-1}$ directly by the hyper-power method. This gives the following sequence of preconditioners:
\begin{equation}
  \label{eq:Pcv}
	\Pc_{V,k+1}^{-1} = 2\Pc_{V,k}^{-1} - \Pc_{V,k}^{-1} \mat A \Pc_{V,k}^{-1}.
\end{equation}

The block $\Pc_{Q,0}$ is a preconditioner for the (negative) Schur complement $\B^T \A^{-1} \B$. To apply the hyper-power method one needs to approximate $\A^{-1}$. Some possibilities are: (1) an approximation by $\Pc^{-1}_{V,0}$, (2) an approximation by a fixed $\Pc^{-1}_{V,{k}}$, (3) an iterative solve by a preconditioned CG method (PCG) with low tolerance and $\Pc^{-1}_{V,0}$, (4) an iterative solve by a PCG method with low tolerance and a fixed $\Pc_{V,k}$,  and (5) an approximation in the $k$th iteration by $\Pc^{-1}_{V,{l}}$, where $l=0$ for $k = 1$ and $l=k$ for $k > 1$. We compare (1) and (5), since (1) has the lowest cost and (5) balances cost and accuracy. We note that the true inverse of the Schur complement is not required; we need merely an approximation of the inverse of $\mat A$. We propose the following sequences of preconditioners for the Schur complement:
\begin{align}
  \label{eq:Pcq1}
	\hat{\Pc}_{Q,k+1}^{-1} &= 2\hat{\Pc}_{Q,k}^{-1} - \hat{\Pc}_{Q,k}^{-1} ( \B^T \Pc^{-1}_{V,l} \B ) \hat{\Pc}_{Q,k}^{-1},\\
  \label{eq:Pcq2}
	\Pc_{Q,k+1}^{-1} &= 2\Pc_{Q,k}^{-1} - \Pc_{Q,k}^{-1} ( \B^T \Pc^{-1}_{V} \B ) \Pc_{Q,k}^{-1},
\end{align}%
which we refer to as Schur complement preconditioners with and without the inner updates, respectively. Note, that $\hat{\Pc}_{Q,0} = \Pc_{Q,0}$ and $\hat{\Pc}_{Q,1} = \Pc_{Q,1}$. For comparison purposes, we also define a sequence where the inverse is computed exactly:
\begin{equation}
  \label{eq:Pcq_exact}
	\bar{\Pc}_{Q,k+1}^{-1} = 2\bar{\Pc}_{Q,k}^{-1} - \bar{\Pc}_{Q,k}^{-1} ( \B^T \A^{-1} \B ) \bar{\Pc}_{Q,k}^{-1}.
\end{equation}

\begin{remark}
  In Equation \eqref{eq:Pcq1} the approximation of the Schur complement depends on $k$, thus the constant $c_A$ in the cost estimate in Equation \eqref{eq:hyperpower_scaling} also depends on $k$.
\end{remark}

\subsection{Preconditioner application cost}
\label{sec:theocost}
The sequences of preconditioners introduced in \eqref{eq:Pcv}, \eqref{eq:Pcq1} and \eqref{eq:Pcq2} involve operators that are block-wise sums of Kronecker product matrices of size $N\times N$. The order of complexity of the matrix-vector product with the $k$th preconditioner in the sequence is
\begin{equation}
  \label{eq:hyperpower_scaling}
  \BigO{C_k N^{4/3}},
\end{equation}
where $C_k = 2^k c_P + (2^k - 1) c_A$ grows exponentially with $k$. Here, $c_P$ is a constant associated with the initial preconditioner, and $c_A$ a constant associated with the approximation. Note that up to the constant $C_k$ the application cost of the $k$th, $k>0$, preconditioner in the sequence scales with the problem size $N$ in the same fashion as the cost of a matrix-vector product with the initial preconditioner. As a consequence, the method is suitable for systems for which the asymptotical cost of the forward problem grows faster with the problem size than that of the initial preconditioner and the approximation. For these systems an increase of the constant $C_k$ can be outweighed by the cost of the forward problem for a growing problem size. We remark that as a consequence of the quadratic convergence of the hyper-power method $k$ is typically small.

%% file: sections/section4_benchmarks.tex
\section{Numerical benchmark}
\label{sec:benchmark}
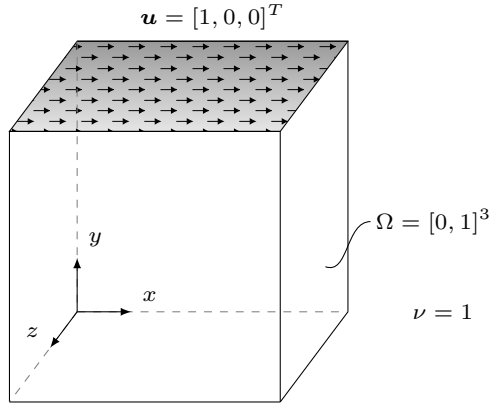
\begin{figure}
    \centering
    \hspace*{7em}
    \begin{tikzpicture}[scale=1.2,transform shape]
      \shade[xslant=0.75, bottom color=gray!20, top color=gray!80]
        (0,0) rectangle +(3,1);
      \fill[xslant=0.75,pattern=arrows] (0,0) rectangle +(3,1);
      \draw[xslant=0.75] (0,0) rectangle +(3,1);
      \draw(0,0) -- +(3,0);
      \draw (0,0) -- (0,-3);
      \draw (3,0) -- (3,-3);
      \draw (3.75,1) -- (3.75,-2);
      \draw[dashed,gray] (0.75,1) -- (0.75,-2);
      \draw[dashed,gray] (0.75,-2) -- (3.75,-2);
      \draw[dashed,gray] (0.75,-2) -- (0,-3);
      \draw (0,-3) -- (3,-3);
      \draw (3.75,-2) -- (3,-3);
  
      \draw[-latex] (0.75,-2) -- ($(0.75,-2)!0.6cm!(0.75,-1.5)$) node[anchor=south west] {\scriptsize{$y$}};
      \draw[-latex] (0.75,-2) -- ($(0.75,-2)!0.6cm!(1.25,-2)$) node[anchor=south west] {\scriptsize{$x$}};
      \draw[-latex] (0.75,-2) -- ($(0.75,-2)!0.5cm!(0,-3)$) node[anchor=south east] {\scriptsize{$z$}};
  
      \draw (3.5,-1.5) to[out=-20,in=180] (4.0,-1);
      \draw (4.7,-1) node {\scriptsize{$\Omega = [0,1]^3$}};
      \draw (4.8,-2) node {\scriptsize{$\nu = 1$}};
      \draw (0.75, 1) -- (3.75,1) node[midway, above] {\scriptsize{$\boldsymbol u = [1,0,0]^T$}};
    \end{tikzpicture}
    \caption{Lid-driven cavity model with unit viscosity, unit velocity at the lid and no-slip boundary conditions on all remaining domain walls.}
    \label{fig:liddrivencavity}
\end{figure}

Let us consider the well-known viscous lid-driven cavity benchmark problem visualized in Figure \ref{fig:liddrivencavity}. We solve the associated linear system of equations resulting from an isogeometric structure-preserving discretization of the Stokes equations. In the following sections we show how matrix-vector products with our Kronecker product operators and preconditioners scale with the problem size. Furthermore, we present the spectra of the preconditioned operators for all sequences of preconditioners introduced in the previous section. The effectiveness of the preconditioners in Equation \eqref{eq:Pcv} and Equation \eqref{eq:Pcq1} is investigated in terms of the number of iterations as well as the solution time of the Minimal Residual method assuming a relative error tolerance of $10^{-8}$.

\subsection{Scaling of matrix-free linear operators}
An important property of the proposed approach is that the hyper-power method does not change the asymptotical scaling of the preconditioner application cost. More precisely, the application of each of the preconditioners in the sequence generated by the hyper-power method using Kronecker product operators scales with $\BigO{N^{4/3}}$.

\begin{figure}
    \centering
    \includegraphics[width=0.95\textwidth]{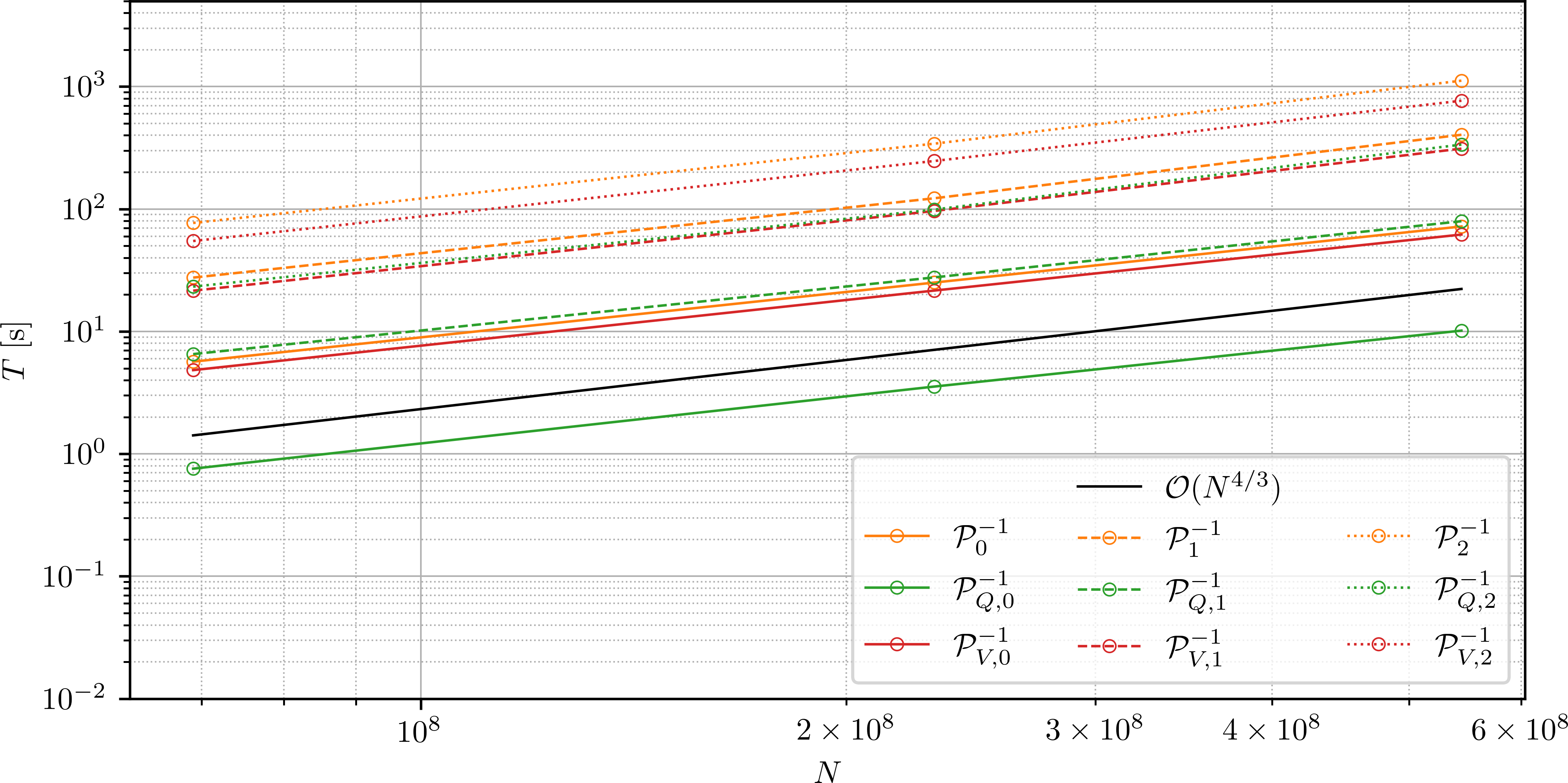}\\
    \vspace*{1em}
    \includegraphics[width=0.95\textwidth]{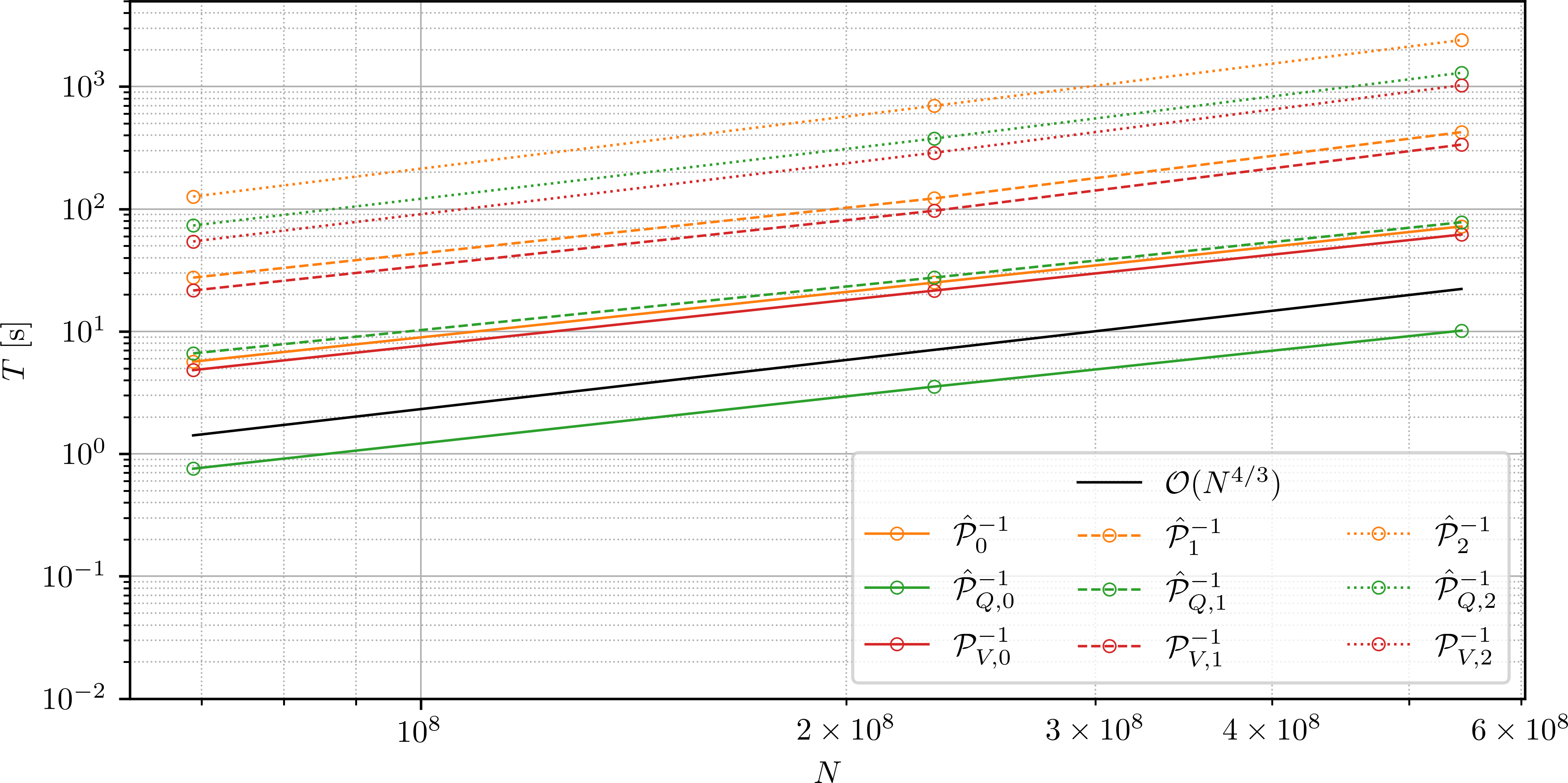}
    \caption{Matrix-vector product benchmarks for the preconditioner sequences defined in Equation~\eqref{eq:Pcq1} (top, without inner updates) and Equation~\eqref{eq:Pcq2} (bottom, with inner updates).}
    \label{fig:timings}
\end{figure}

In Figure \ref{fig:timings} we present benchmarks for matrix-vector products with all matrix-free linear operators involved in the solution of the model problem on meshes with $256^3$, $384^3$ and $512^3$ elements. Our benchmarks verify that matrix-vector products with each of the operators scale with $N^{4/3}$ times a constant, where the constant depends on the operator in question. \emph{All timings were obtained on a single thread.}

We now apply the cost estimate of Equation \eqref{eq:hyperpower_scaling} to the sequence in Equation~\eqref{eq:Pcv}. In our implementation the cost constants for the fast diagonalization preconditioner and the operator $\A$ are $c_P = 6\cdot 3^{-4/3}$ and $c_A = 15\cdot 3^{-4/3}$. Estimate \eqref{eq:hyperpower_scaling} predicts that the cost of applying $\hat{\Pc}_{V,1}$ and $\hat{\Pc}_{V,2}$ is $4.5$ and $11.52$ times the cost of applying $\hat{\Pc}_{V,0}$, respectively. In our benchmark a matrix-vector product with $\Pc_{V,0}$ on the mesh with $384^3$ elements takes $21.58$ seconds. A matrix-vector product with  $\hat{\Pc}_{V,1}$ and $\hat{\Pc}_{V,2}$ takes $96.44$ and $246.82$ seconds, respectively. These timings are in agreement with the estimate of $97.11$ and $248.60$ seconds. Similar estimates can be obtained for the remaining operators, but are omitted here for brevity.

\subsection{Spectra of preconditioned operators}
The spectra discussed in this section correspond to a discretization of degree $(p_1, p_2, p_3) = (4,4,4)$ and maximum regularity. The mesh is uniform and consists of $2\times 2\times 2$ elements.
\begin{figure}
    \centering
    \includegraphics[width=0.95\textwidth]{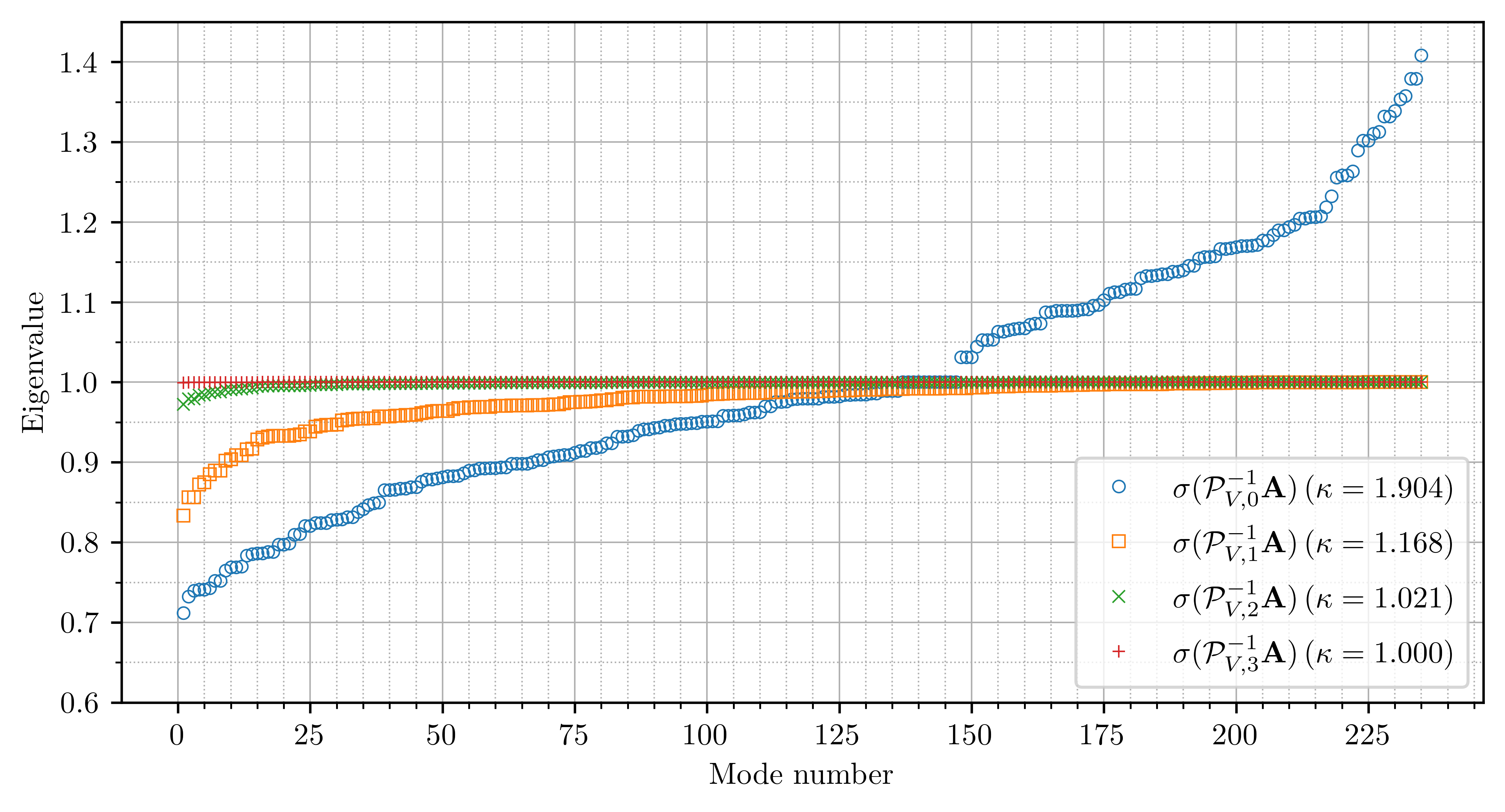}
    \caption{Spectra and condition numbers of the preconditioned operator $\A$ for the sequence of preconditioners
    in Equation \eqref{eq:Pcv}.}
    \label{fig:spectrum_opA}
\end{figure}

In Figure~\ref{fig:spectrum_opA} we present the spectra of the operator $\A$ preconditioned by the sequence of preconditioners in Equation \eqref{eq:Pcv}. In this sequence we use the exact operator $\mat A$, thus we can use the spectra to review the theoretical results in Section \ref{sec:preconditioning_strategy}. We find $\lambda_{0,{\rm min}}=0.71$ and $\lambda_{0,{\rm max}}=1.41$ which confirms the assumption $\sigma(\PinvA_0) \subset (0,2)$. Hence, from Remark~\ref{rmk:convergence_Pk} we observe that $\mathcal P_k$ in Equation \eqref{eq:Pcv} converges and Theorem~\ref{thm:positive definite} ensures that $\mathcal P_k$, $k \geq 0$, is positive definite. Additionally, Corollary~\ref{cor:Skform} guarantees $\lambda_{k+1,{\rm max}} \leq 1$, $k \geq 0$. Furthermore, starting from the smallest and largest eigenvalue in the blue data set, Corollary~\ref{cor:Skform} yields $\lambda_{k,{\rm min}} = \List{ 0.71, 0.8319, 0.9717, 0.9992 }$, $k=0,1,2,3$, respectively. These eigenvalues can be readily verified in Figure~\ref{fig:spectrum_opA}. Note that we obtain an almost exact inverse after just three updates of the hyper-power method. In other words, only three updates are necessary to include the effects of the off-diagonal blocks of $\mat A$ in the fast diagonalization preconditioner. Finally, we observe decreasing condition numbers $\kappa(\PinvA_{k})$ with each update. This is in correspondence with Corollary~\ref{cor:decreasing_cond_numbers}.

\begin{figure}
    \centering
    \includegraphics[width=0.495\textwidth]{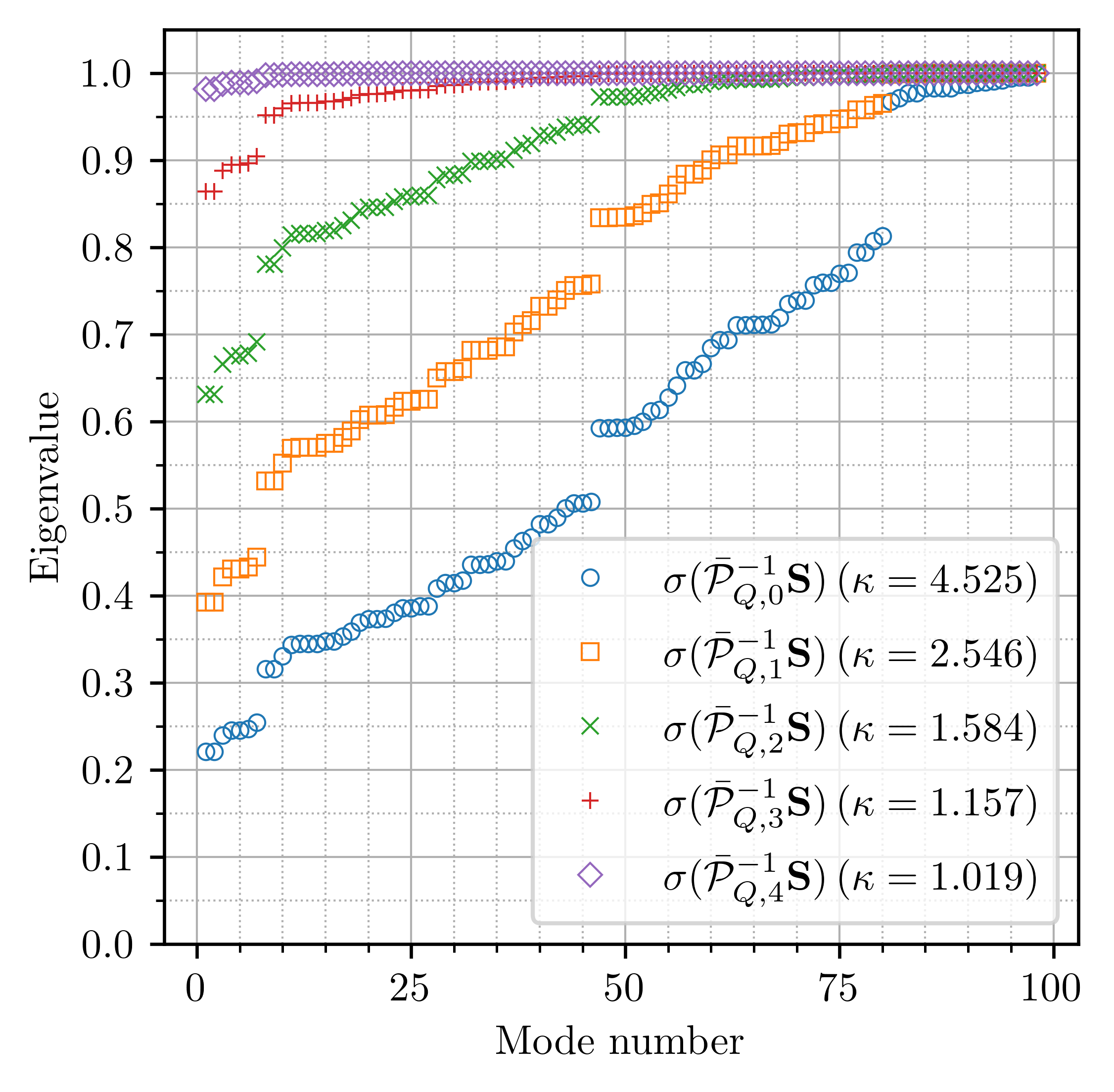}
    \includegraphics[width=0.495\textwidth]{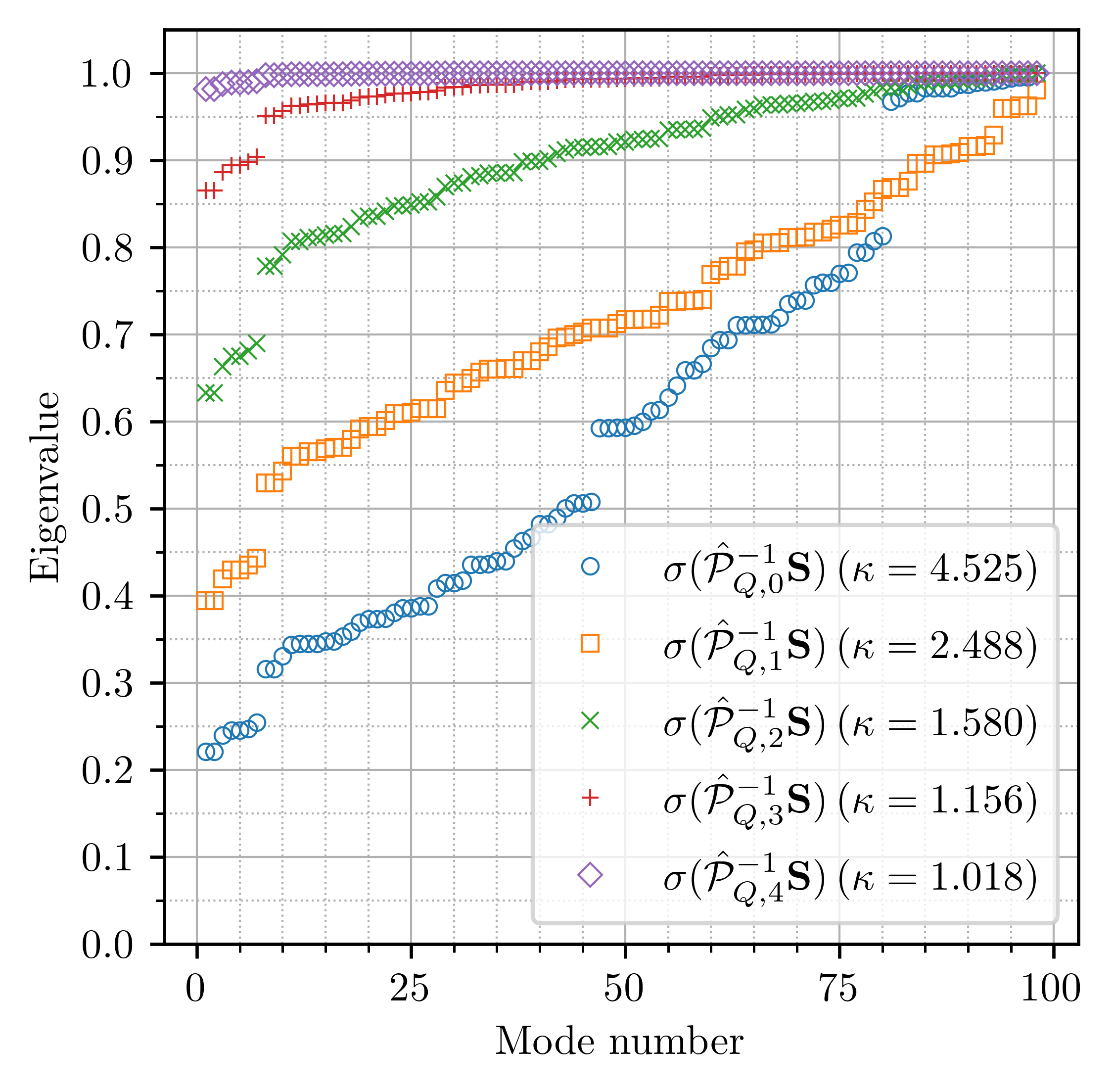}
    \includegraphics[width=0.90\textwidth]{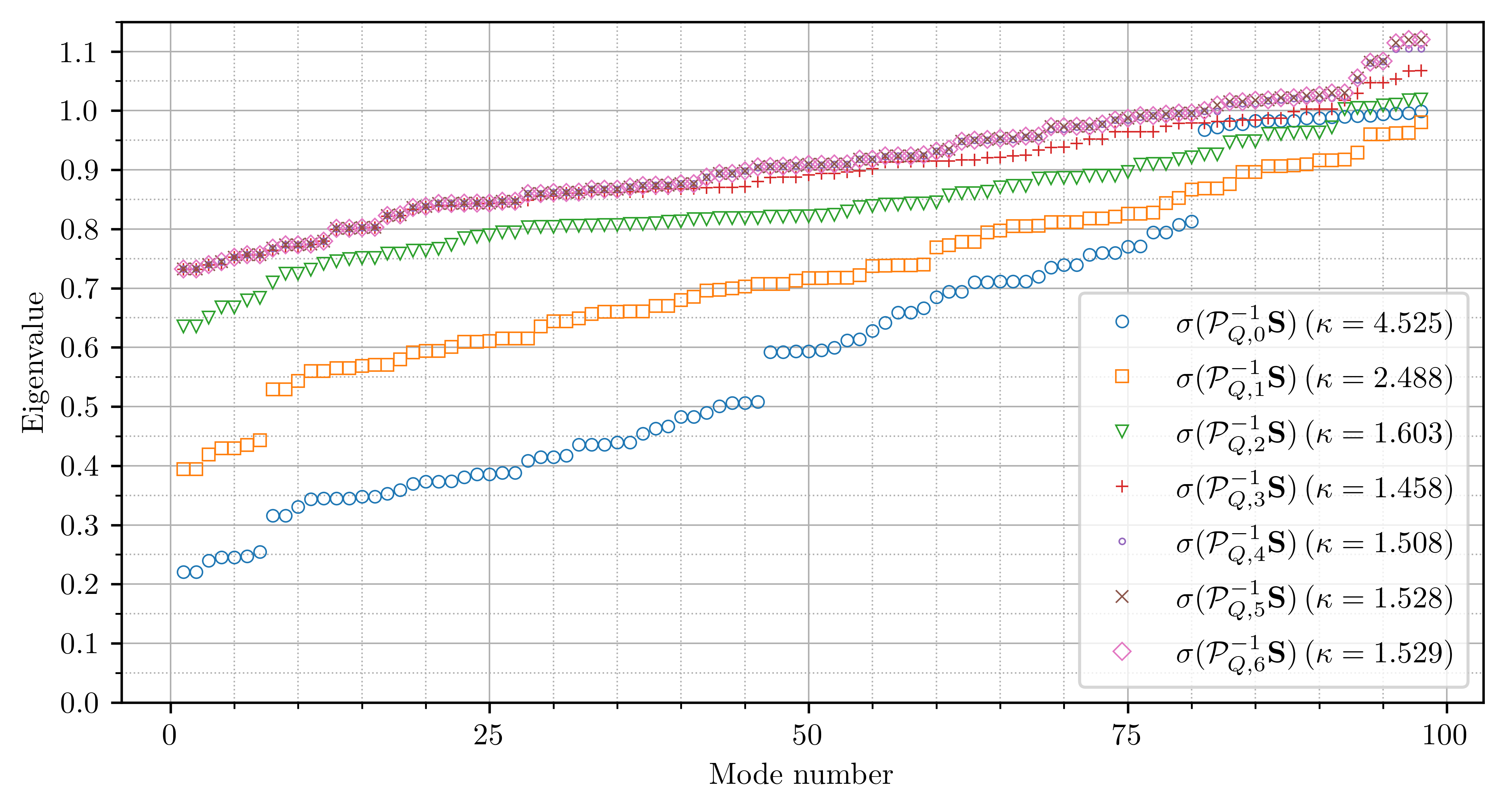}
    \caption{Spectra and condition numbers of the preconditioned Schur complement for the sequence of preconditioners
    in Equation \eqref{eq:Pcq_exact} with $\A^{-1}$ exact (top-left), the sequence in Equation \eqref{eq:Pcq1} with inner updates (top-right) and the sequence in Equation \eqref{eq:Pcq2} without inner updates (bottom).}
    \label{fig:Pcq_exact_and_updated}
\end{figure}
Next, we consider the preconditioners for the Schur complement. The spectra corresponding to the sequences in Equations~\eqref{eq:Pcq1} and \eqref{eq:Pcq_exact} are shown in the top row of Figure \ref{fig:Pcq_exact_and_updated}. The spectra on left behave analogously to the spectra discussed previously and the improvement of the condition number is more significant. The spectra on the left and right are almost indistinguishable. In particular, the upper bound of the spectrum is around the value of one. Surprisingly, the updated approximation, Equation \eqref{eq:Pcq1}, performs slightly better than the exact inverse of $\A^{-1}$, Equation \eqref{eq:Pcq_exact}, which is reflected in slightly lower condition numbers.

The spectra in the bottom row of Figure \ref{fig:Pcq_exact_and_updated} correspond to the sequence of Schur complement preconditioners in Equation \eqref{eq:Pcq2}. These preconditioners are particularly inexpensive, because the inverse of $\A$ is approximated by fast diagonalization. Due to this approximation, the hyper-power method converges to an approximation of the inverse of the Schur complement, which ignores the off-diagonal blocks of the operator~$\A$. Thus, this sequence of preconditioners is not expected to perform superior to the sequences discussed previously. In fact, only the first three iterations lead to an improvement of the condition number. Ignoring the off-diagonal blocks might be justifiable, considering the low cost and the fact that the first three updates perform well.

\subsection{Preconditioner effectiveness}

In Table~\ref{tab:liddrivencavity}, we present the iteration numbers versus the normalized solution time for mesh sizes $1/8$, $1/16$, $1/32$ and $1/64$, as well as polynomial degrees $2$, $4$, $6$ and~$8$. Based on the spectra in Figure \ref{fig:spectrum_opA} and \ref{fig:Pcq_exact_and_updated}, we restrict the study to four updates of the hyper-power method in Equations~\eqref{eq:Pcv}~and~\eqref{eq:Pcq1}. Further reduction of the number of iterations is limited by the block-diagonal structure of the preconditioner in Equation \eqref{eq:blockdiagprecond} rather than the number of updates. To evaluate the forward problem, we use matrix-free formation and assembly in combination with global sum factorization and Gauss--Legendre quadrature with $(p+1)$ points in each univariate direction \cite{BrTa2018}. The solution time is normalized with respect to the solution time of the system preconditioned by the initial preconditioner. In Figure~\ref{fig:Pc_efficiency_benchmark1}, we visualize the cases using $\Pc_k$, $k=0,1,\ldots,4$, for mesh sizes $1/16$ and $1/64$ and degrees $2$, $4$, $6$ and $8$ in a contour plot first introduced in Section \ref{sec:introduction}. In contrast to Figure~\ref{fig:Pc_efficiency}, the number of iterations and the solution time are normalized by the solution time and the number of iterations of the system preconditioned by the initial preconditioner. As indicated by the spectra, the number of iterations is robustly reduced with each update. A single update of the hyper-power method reduces the solution time by roughly a half in all test cases. The number of iterations and the solution time reduce further with decreasing mesh size, increasing polynomial order and number of updates. The solution time is reduced by $86\%$ in the case with $64^3$ elements, a polynomial order of $8$, and $4$ updates. Note that the results depend on the cost of the forward problem. We remark that there are more efficient quadrature techniques \cite{antolin2015efficient,calabro2017fast,HiSaTaCaHu2019}. For the presented benchmark on a regular grid a quadrature-free approach in conjunction with Kronecker products is the fastest approach, but is not relevant for more general applications.

\begin{table}[h]
    \centering
    \caption{Performance of the Minimal Residual method in the lid-driven cavity benchmark in terms of the number of iterations and the solution time normalized by the solution time of the system preconditioned by the initial preconditioner.}
    \label{tab:liddrivencavity}
    \renewcommand{\arraystretch}{1.25}
    \resizebox{\columnwidth}{!}{%
    \begin{tabular}{!{\vrule width 1pt}c!{\vrule width 1pt}c!{\vrule width 1pt}c|c|c|c|c|c|c|c|c|c!{\vrule width 1pt}}
    \noalign{\hrule height 1pt}
        \multirow{2}{*}{Mesh}             & \multirow{2}{*}{Degree} & \multicolumn{2}{c|}{$\hat{\mathcal P}_0$} & \multicolumn{2}{c|}{$\hat{\mathcal P}_1$} & \multicolumn{2}{c|}{$\hat{\mathcal P}_2$} & \multicolumn{2}{c|}{$\hat{\mathcal P}_3$} & \multicolumn{2}{c!{\vrule width 1pt}}{$\hat{\mathcal P}_4$} \\    \cline{3-12}
                                          &                         & $N_{\rm iter}$ & $T_{\rm sol}$            & $N_{\rm iter}$ & $T_{\rm sol}$            & $N_{\rm iter}$ & $T_{\rm sol}$            & $N_{\rm iter}$ & $T_{\rm sol}$            & $N_{\rm iter}$ & $T_{\rm sol}$            \\    \noalign{\hrule height 1pt}
        \multirow{5}{*}{\centering $8^3$} & $2$                     & $63$ & $1.0$ & $36$ & $0.61$ & $24$ & $0.51$ & $17$ & $0.56$ & $11$ & $0.71$ \\
                                          & $4$                     & $59$ & $1.0$ & $32$ & $0.54$ & $21$ & $0.39$ & $14$ & $0.33$ & $9$ & $0.32$ \\ 
                                          & $6$                     & $54$ & $1.0$ & $29$ & $0.54$ & $18$ & $0.35$ & $12$ & $0.26$ & $7$ & $0.19$ \\ 
                                          & $8$                     & $56$ & $1.0$ & $29$ & $0.52$ & $19$ & $0.35$ & $12$ & $0.24$ & $7$ & $0.16$ \\ \hline
        \multirow{5}{*}{\centering $16^3$}& $2$                     & $66$ & $1.0$ & $36$ & $0.58$ & $24$ & $0.46$ & $16$ & $0.43$ & $11$ & $0.52$ \\
                                          & $4$                     & $56$ & $1.0$ & $30$ & $0.54$ & $20$ & $0.38$ & $14$ & $0.30$ & $9$ & $0.25$ \\ 
                                          & $6$                     & $54$ & $1.0$ & $28$ & $0.52$ & $18$ & $0.35$ & $12$ & $0.24$ & $7$ & $0.16$ \\ 
                                          & $8$                     & $54$ & $1.0$ & $29$ & $0.54$ & $19$ & $0.36$ & $12$ & $0.23$ & $7$ & $0.15$ \\ \hline
        \multirow{5}{*}{\centering $32^3$}& $2$                     & $66$ & $1.0$ & $36$ & $0.57$ & $24$ & $0.46$ & $16$ & $0.43$ & $11$ & $0.51$ \\
                                          & $4$                     & $56$ & $1.0$ & $30$ & $0.54$ & $20$ & $0.38$ & $14$ & $0.30$ & $9$ & $0.25$ \\ 
                                          & $6$                     & $52$ & $1.0$ & $28$ & $0.54$ & $18$ & $0.35$ & $12$ & $0.25$ & $7$ & $0.16$ \\ 
                                          & $8$                     & $54$ & $1.0$ & $28$ & $0.52$ & $18$ & $0.34$ & $12$ & $0.23$ & $7$ & $0.14$ \\ \hline
        \multirow{5}{*}{\centering $64^3$}& $2$                     & $66$ & $1.0$ & $34$ & $0.53$ & $24$ & $0.42$ & $16$ & $0.37$ & $11$ & $0.41$ \\
                                          & $4$                     & $56$ & $1.0$ & $30$ & $0.54$ & $20$ & $0.37$ & $14$ & $0.28$ & $9$ & $0.21$ \\ 
                                          & $6$                     & $52$ & $1.0$ & $28$ & $0.54$ & $18$ & $0.35$ & $12$ & $0.24$ & $7$ & $0.15$ \\ 
                                          & $8$                     & $53$ & $1.0$ & $28$ & $0.53$ & $18$ & $0.34$ & $12$ & $0.23$ & $7$ & $0.14$ \\  
    \noalign{\hrule height 1pt}
    \end{tabular}%
    }
\end{table}

\begin{figure}
    \begin{tabular}{cc}
        \includegraphics[width=0.45\textwidth]{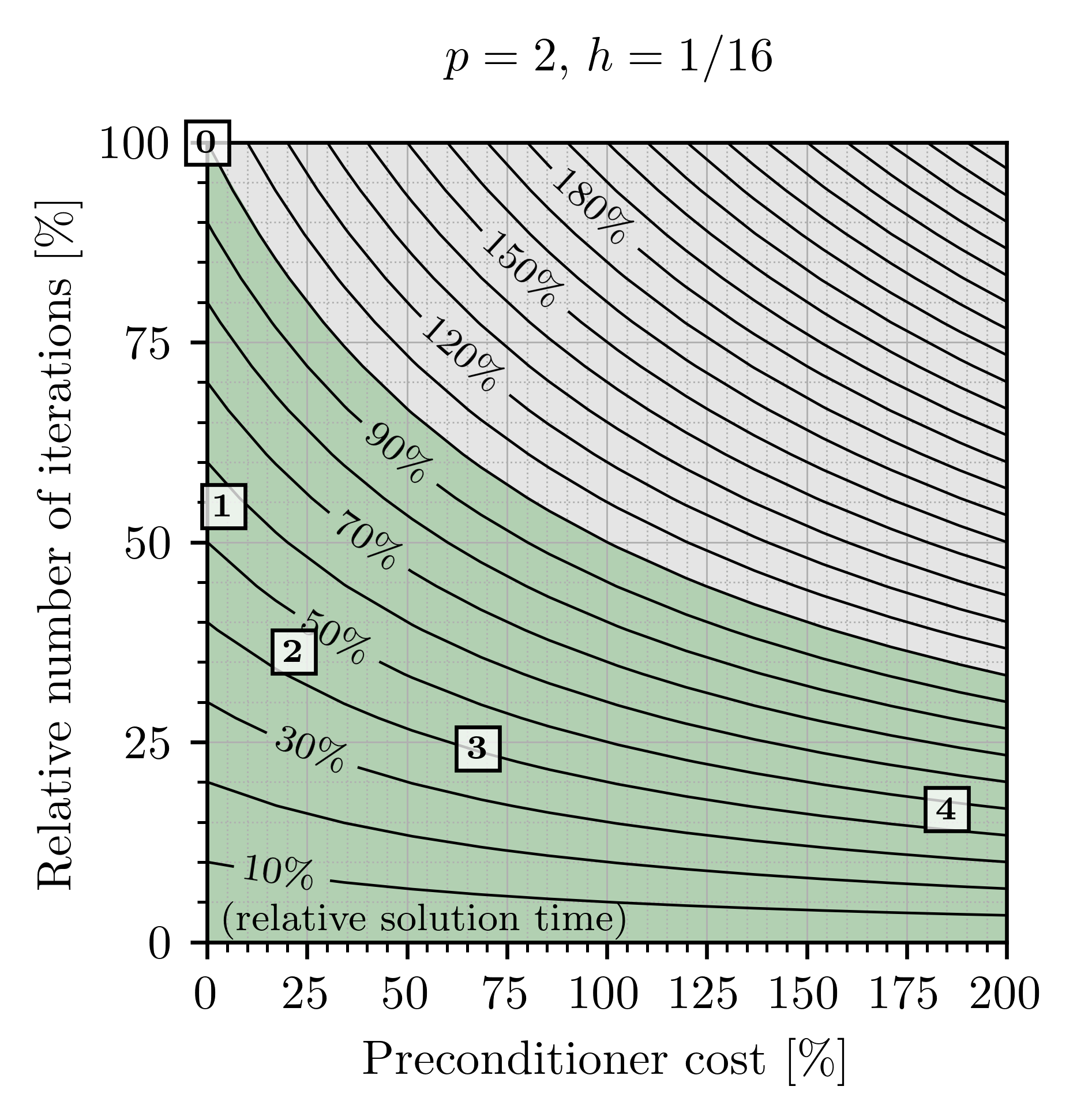} & \includegraphics[width=0.45\textwidth]{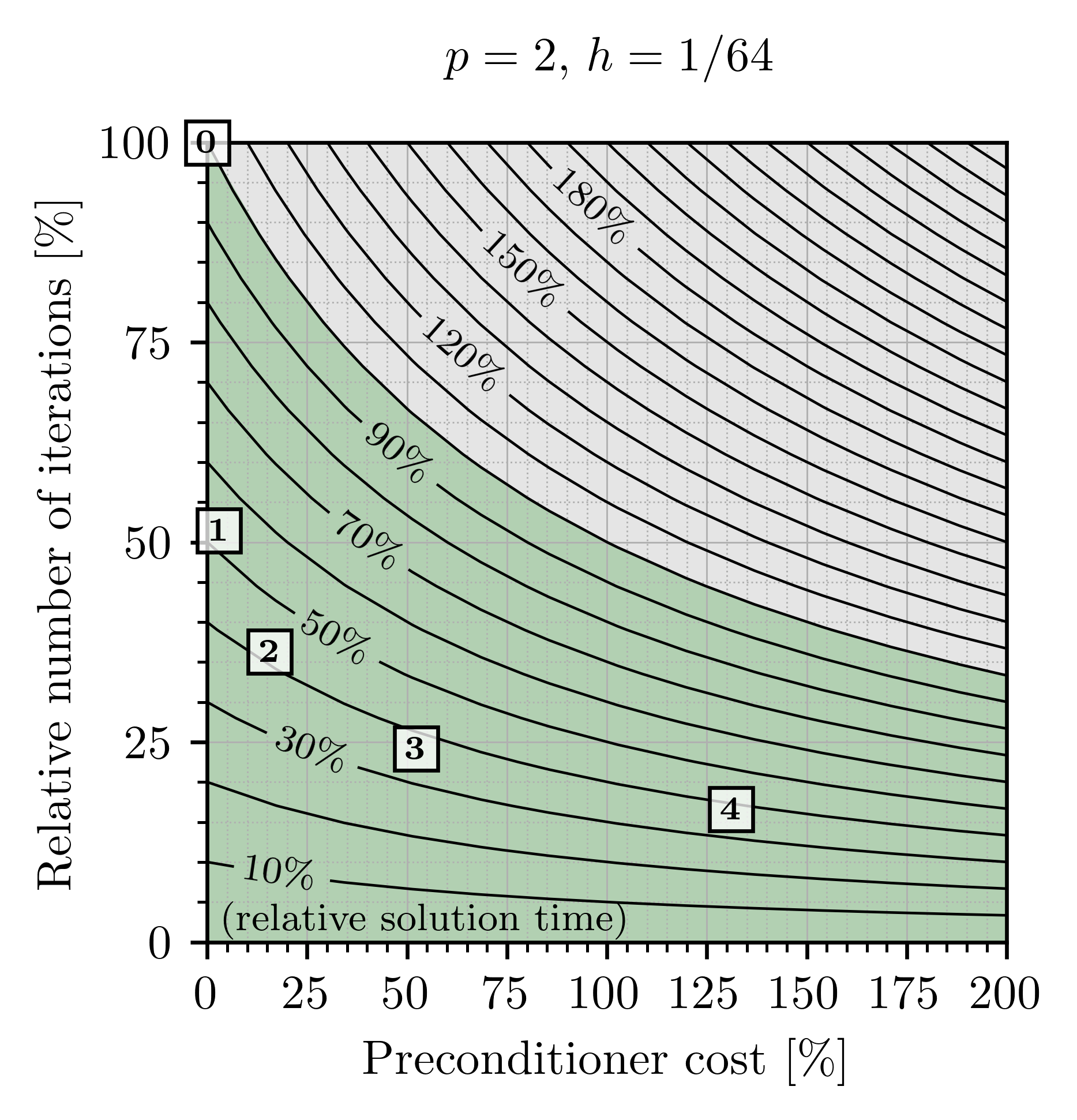}
    \end{tabular}
    \caption{Contour plots for solution time ratio in the lid-driven cavity benchmark in terms of preconditioner cost ratio and number of iterations ratio. The ratios are computed with respect to the benchmark case with the initial preconditioner. Labels attached to data points denote the index~$k$ of the corresponding preconditioner $\hat{\Pc}_k$.}
    \label{fig:Pc_efficiency_benchmark1}
\end{figure}
\begin{figure}
    \ContinuedFloat
    \captionsetup{list=off,format=cont}
    \begin{tabular}{cc}
        \includegraphics[width=0.45\textwidth]{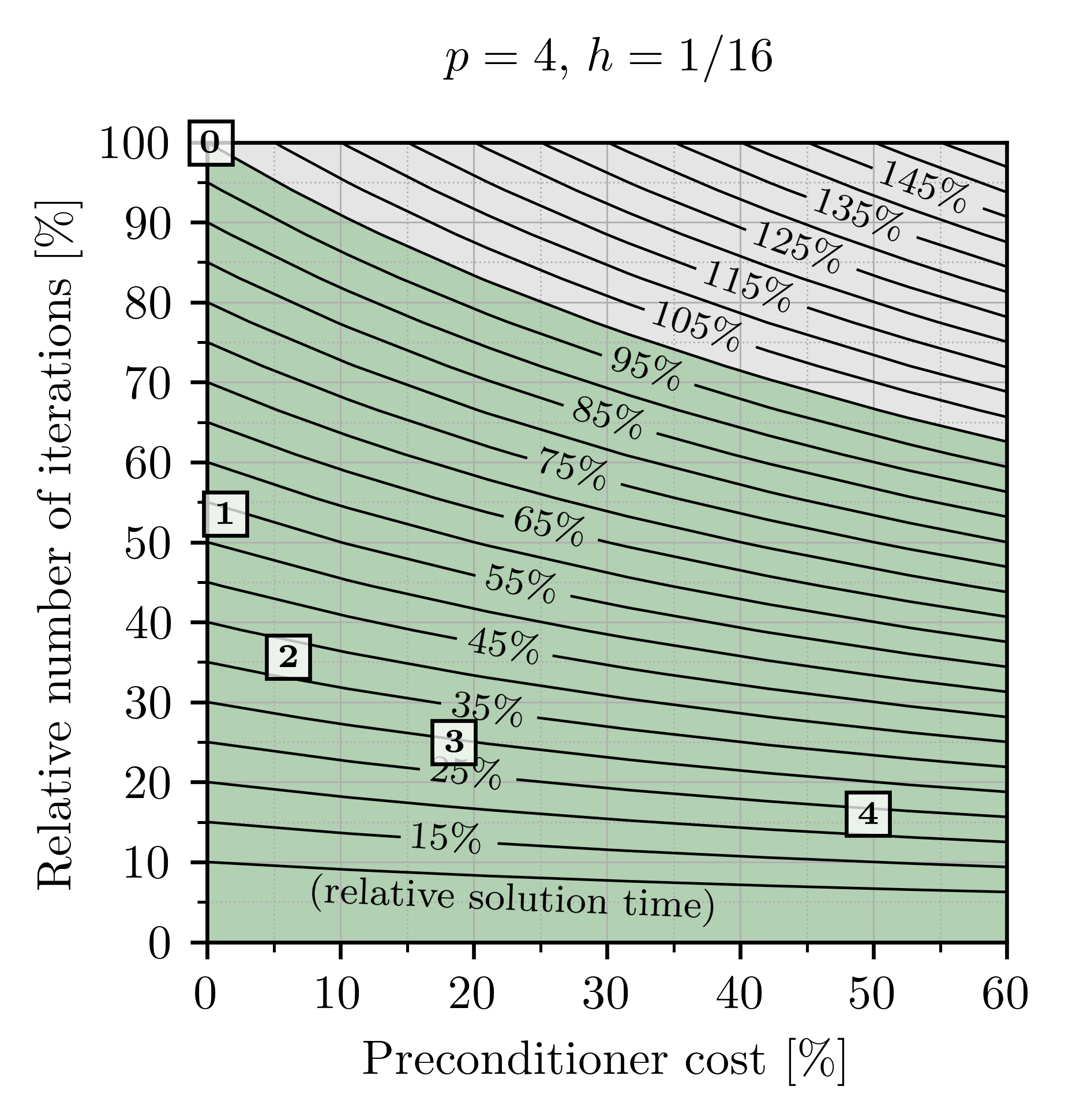} & \includegraphics[width=0.45\textwidth]{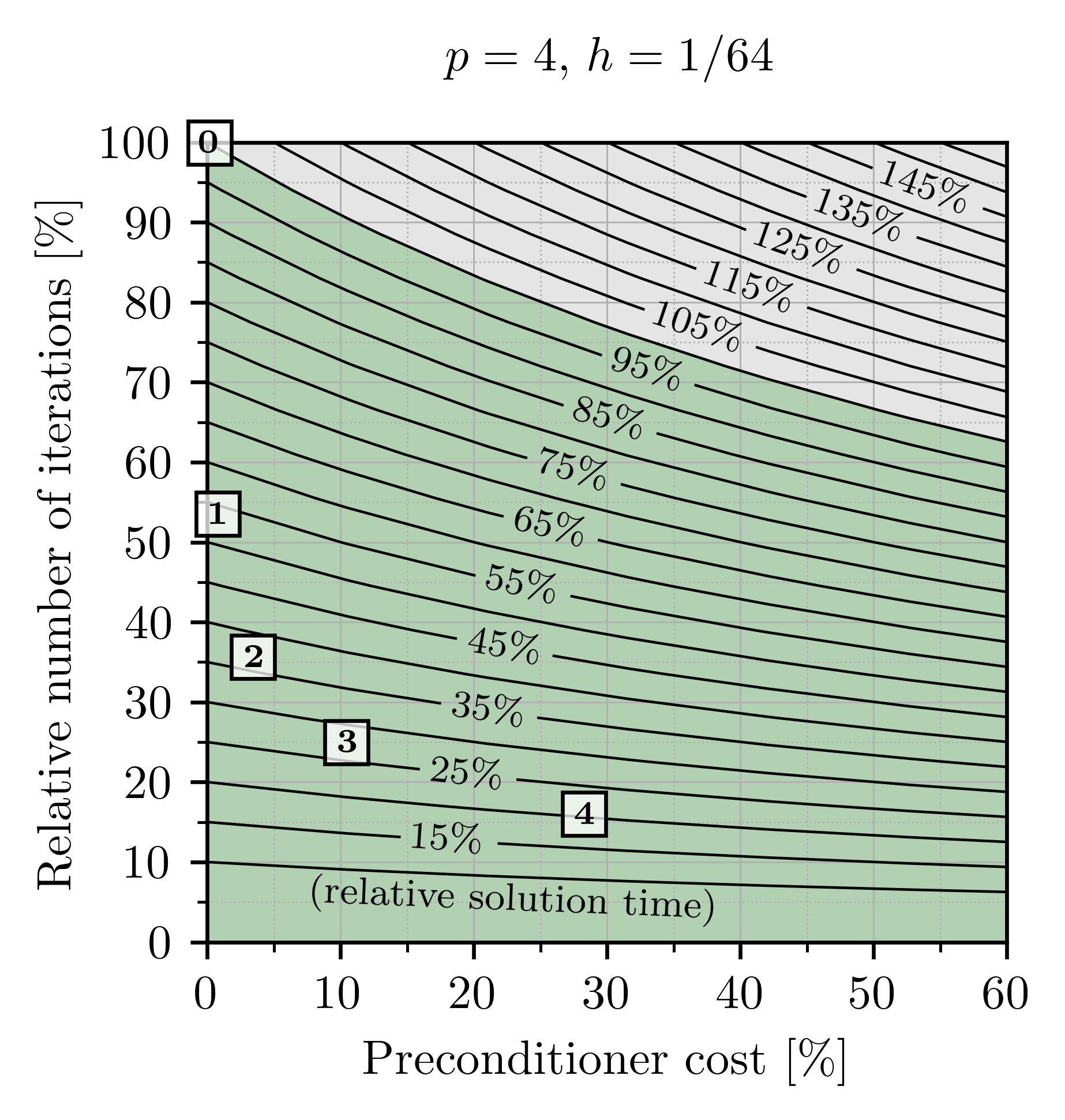} \\
        \includegraphics[width=0.45\textwidth]{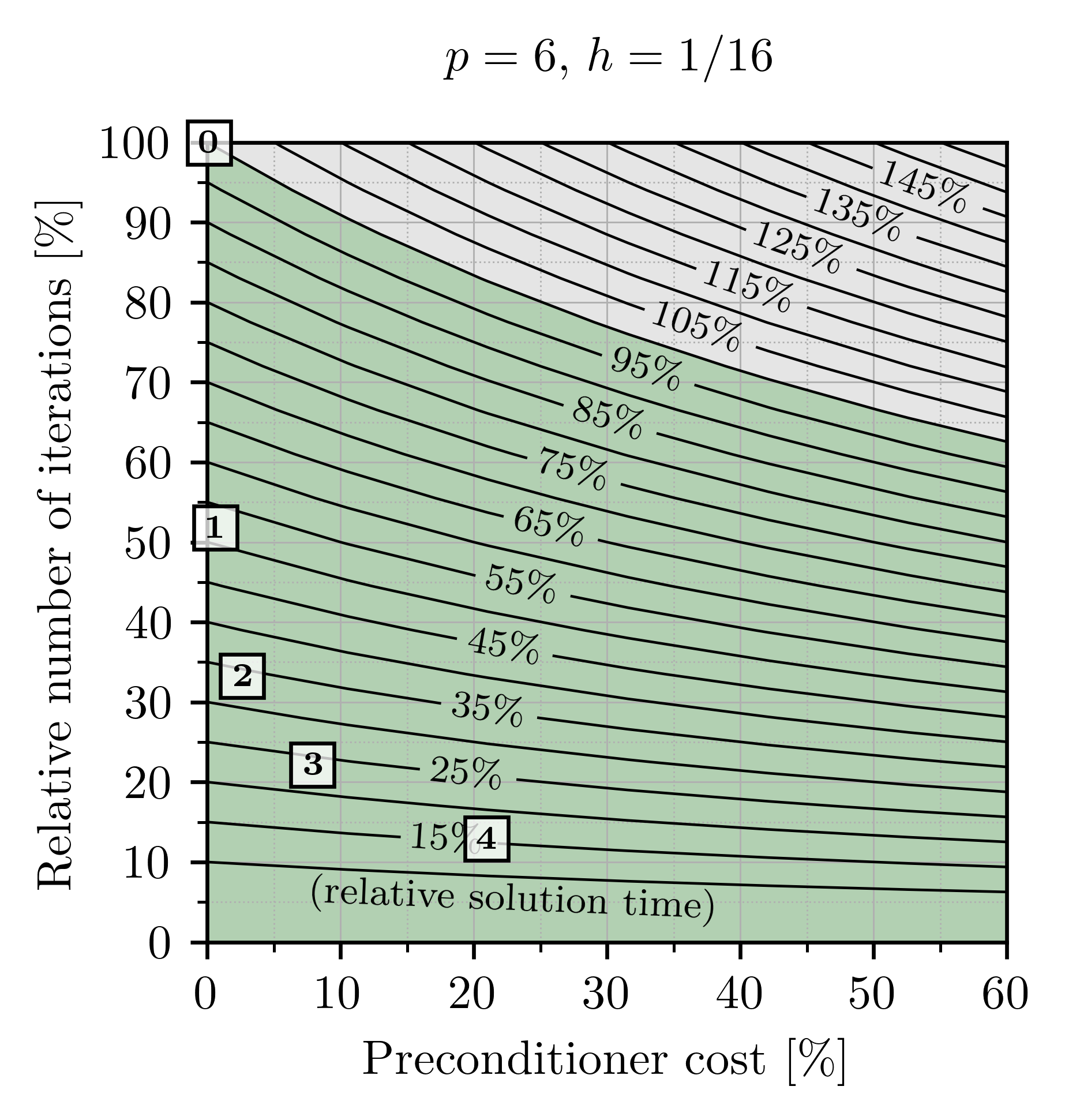} & \includegraphics[width=0.45\textwidth]{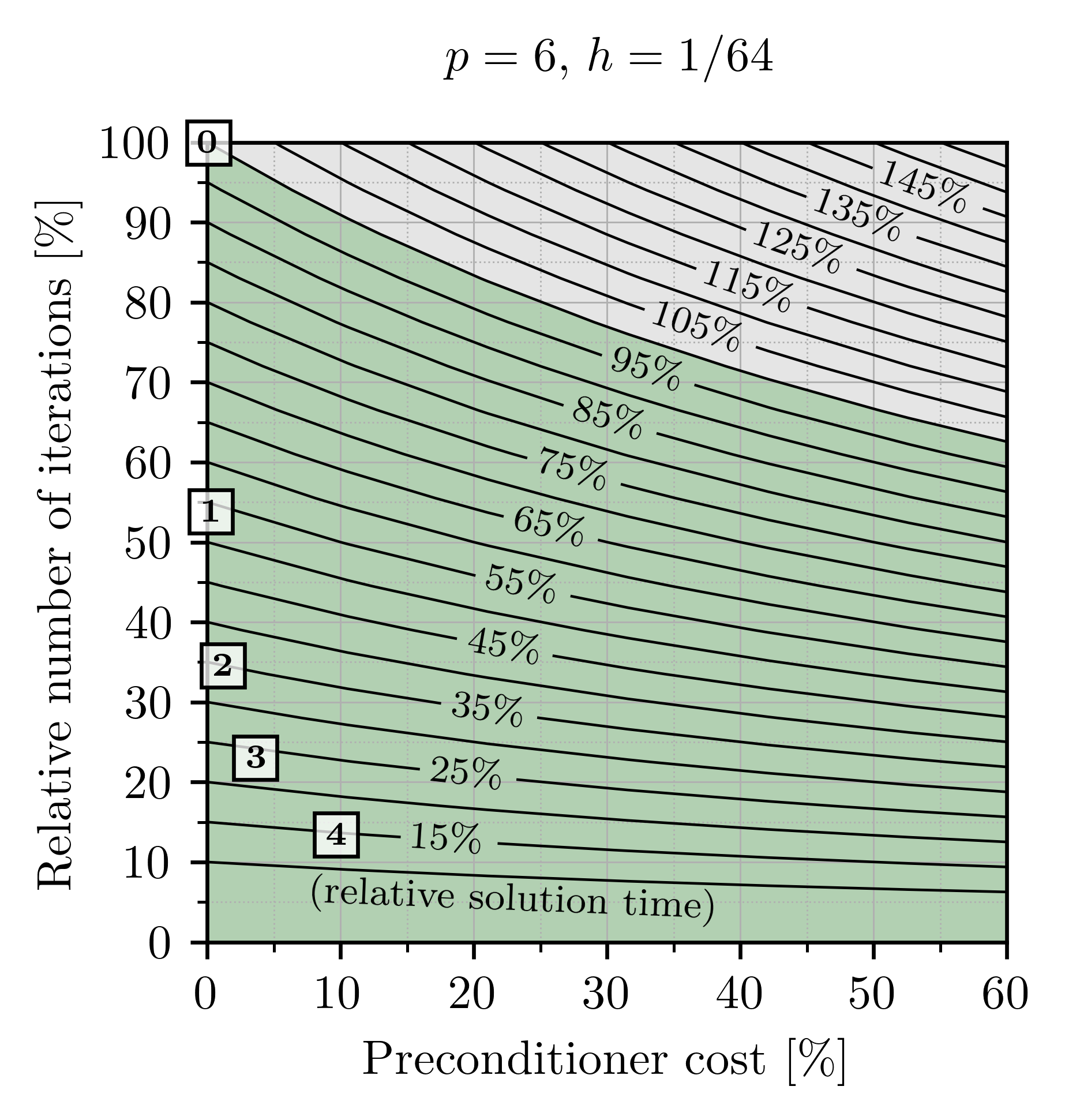} \\
        \includegraphics[width=0.45\textwidth]{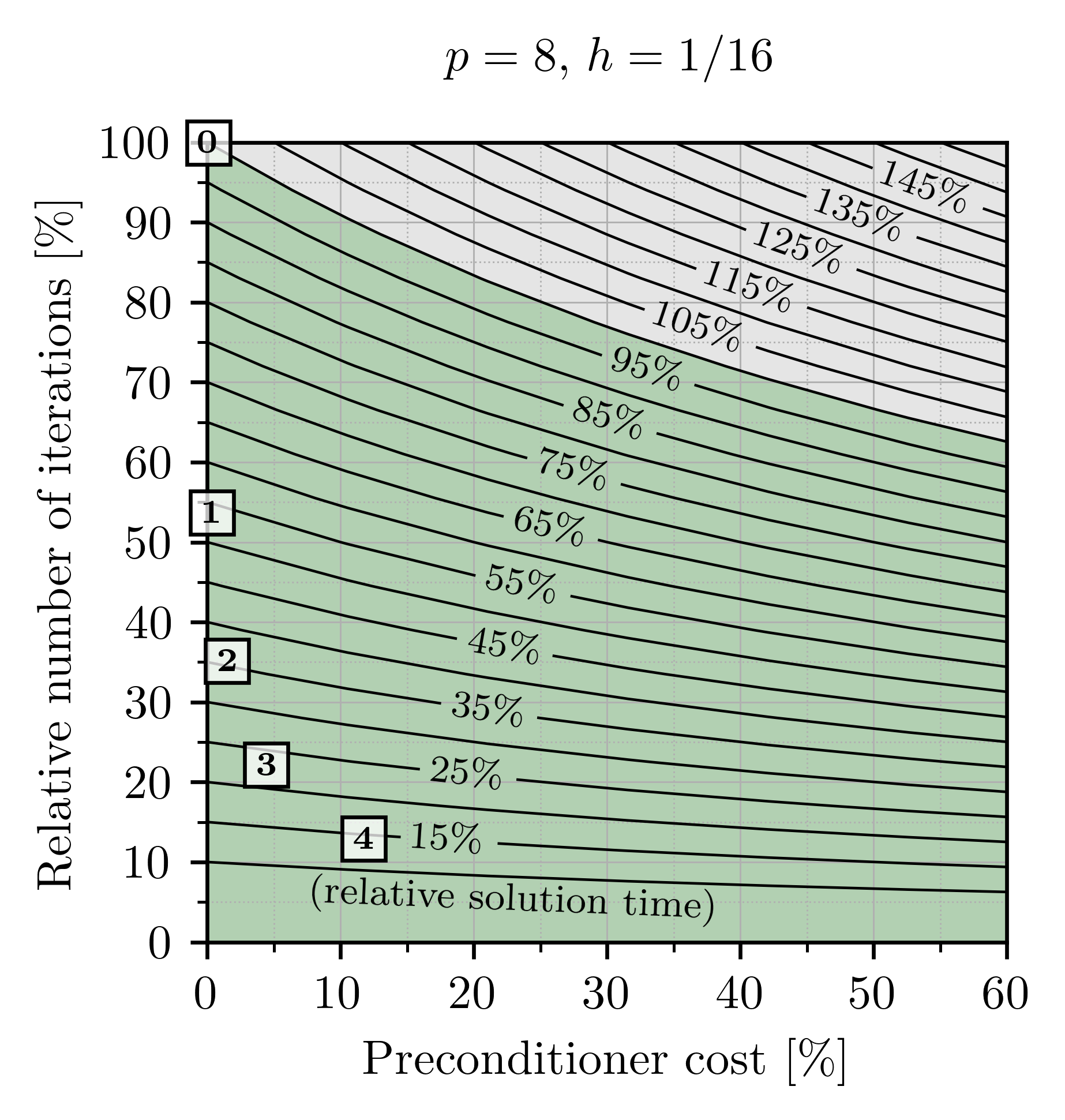} & \includegraphics[width=0.45\textwidth]{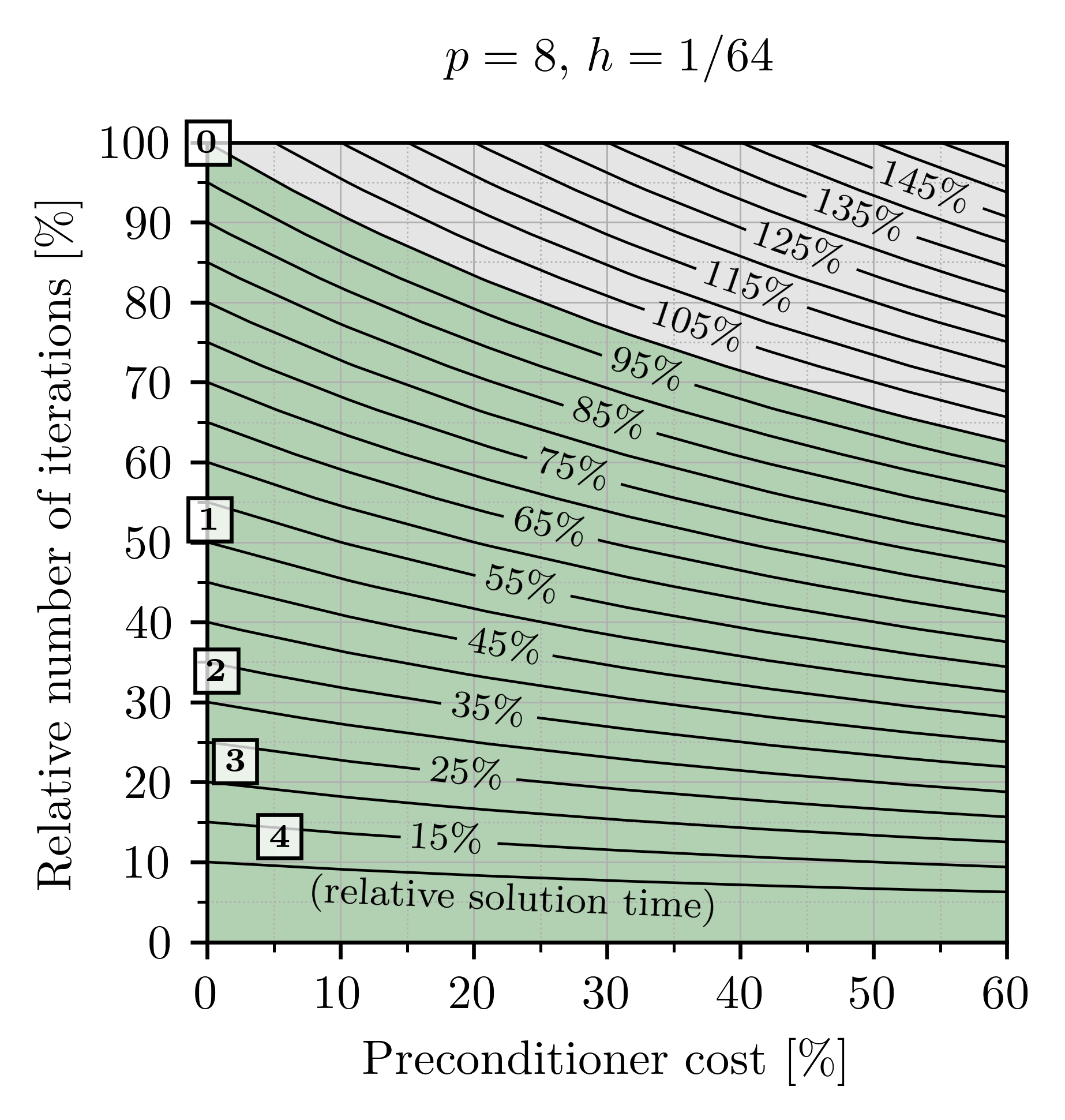} 
    \end{tabular}
    \caption{Contour plots for solution time ratio in the lid-driven cavity benchmark in terms of preconditioner cost ratio and number of iterations ratio. The ratios are computed with respect to the benchmark case with the initial preconditioner. Labels attached to data points denote the index~$k$ of the corresponding preconditioner $\hat{\Pc}_k$.}
\end{figure}

%% file: sections/section5_conclusion.tex
\section{Conclusion} \label{sec:conclusion}
In this paper, we introduce matrix-free polynomial preconditioners based on Schultz's second-order hyper-power method, a self-correcting method of successive approximation of an inverse matrix. Our focus is to solve a saddle point system using MINRES, derived from isogeometric structure preserving discretization of the Stokes equations. We initiate the sequence with a state-of-the-art block-diagonal preconditioner and apply $k$ steps of the hyper-power method, which corresponds to a truncated Neumann series of order $2^k$.

While polynomial preconditioners cannot reduce the total matrix-vector multiplications in Krylov methods, our approach significantly cuts costs by using precise approximations of the matrix operators via Kronecker products, leading to highly efficient matrix-vector products. Eigenvalue analysis guides the selection of the initial preconditioner for convergence. Our results show that multiple iterations of the hyper-power method, along with accurate matrix operator approximations, substantially reduce iteration counts and overall solution time, particularly in simulating Stokes flow inside a 3D lid-driven cavity.

Looking ahead, we plan to extend this method to the incompressible Navier-Stokes equations, which involves skew-symmetric operators. Here, an iterative Krylov method like GMRES could benefit greatly from a more effective (and more expensive to apply) preconditioner since its cost scales superlinear with iteration count. Additionally, we aim to combine our technique with multigrid methods.

%% file: sections/appendix1_diagonalization.tex
\section{The Kronecker product}
\label{app:kronecker}

\subsection{Kronecker products and their properties}
The properties of Kronecker product matrices are essential in the design of the operators and preconditioners introduced in Sections \ref{sec:model} and \ref{sec:preconditioning_strategy}. Let $\mmat{A} \in \mathbb{R}^{m \times n}$, $\mmat{B} \in \mathbb{R}^{p \times q}$, and $\mmat{C} \in \mathbb{R}^{s \times t}$ denote real valued matrices. The Kronecker product $\mmat{A} \otimes \mmat{B} \in \mathbb{R}^{m \cdot p \times n \cdot q}$ is the matrix
\begin{align}
	\mmat{A} \otimes \mmat{B} := 
	\begin{bmatrix}
		A_{11} \mmat{B} 	& \ldots  	& A_{1n}	\mmat{B} 	\\
		\vdots 					&				& 	\vdots 		\\
		A_{m1} \mmat{B} 	& \ldots  	& A_{mn}	\mmat{B}
	\end{bmatrix}.
\end{align}
Kronecker products satisfy the following properties:
\begin{align}
	\left( \mmat{A} \otimes \mmat{B}   \right) \otimes \mmat{C} &= \mmat{A} \otimes \left( \mmat{B} \otimes \mmat{C}   \right)	&& (\text{associativity})	\\
	\left( \mmat{A} \otimes \mmat{B}   \right) \left( \mmat{C} \otimes \mmat{D}   \right)  &= \left( \mmat{A} \mmat{C} \right) \otimes \left( \mmat{B} \mmat{D}   \right)	&& (\text{mixed product property})	\\
	\left( \mmat{A} \otimes \mmat{B}   \right)^{-1} &= \mmat{A}^{-1} \otimes \mmat{B}^{-1}	&& (\text{inverse of a Kronecker product})\\
	\left( \mmat{A} \otimes \mmat{B}   \right)^{T} &= \mmat{A}^{T} \otimes \mmat{B}^{T} && (\text{transpose of a Kronecker product}).
\end{align}

\subsection{Kronecker matrix-vector multiplication}
Let $\mmat{X} \in \mathbb{R}^{n_1 \times \cdots \times n_d}$ and $\mmat{Y} \in \mathbb{R}^{m_1 \times \cdots \times m_d}$ denote two $d$-dimensional arrays. Vectorization of $\mmat{X}$ is a linear operation that maps $\mmat{X}$ to a vector  $\text{vec}(\mmat{X}) \in \mathbb{R}^{n_1 \cdot \ldots \cdot n_d}$ with entries
\begin{align}
	\text{vec}(\mmat{X})_i := X_{i_1 \ldots i_d}, \text{ where } i=i_1 + (i_2-1)n_1 + \cdots + (i_d-1) n_1 \cdot \ldots \cdot n_{d-1}.
\end{align}
One recurring theme in this paper involving Kronecker matrices is efficient matrix-vector multiplication. Let $\mmat{D}_k \in \mathbb{R}^{m_k \times \n_k}$ denote a set of $d$ matrices $\List{D_{i_k \; j_k} , \; i_k \in 1, \ldots , m_k, \; j_k \in 1, \ldots , n_k,\; \text{for } k=1, \ldots , d}$. The matrix vector product
\begin{align}
	\text{vec}(\mmat{Y}) &= \left( \mmat{D}_d \otimes \ldots \otimes \mmat{D}_1  \right) \text{vec}(\mmat{Y}) &&
	\BigO{M \cdot N} \text{ flops}
\end{align}
can be written as a tensor contraction instead,
\begin{align}
	Y_{j_1 \ldots j_d} &= \sum_{j_1 \ldots j_d} D_{i_1 j_1} \ldots D_{i_d j_d} \;  X_{i_1 \ldots i_d} &&
	\BigO{\max{(N \cdot m_1, \; n_d \cdot M)}} \text{ flops}.
\end{align}
Here $N= n_1 \cdot \ldots \cdot n_d$ and $M=m_1 \cdot \ldots \cdot m_d$. The second approach scales nearly linearly with matrix size and significantly outperforms standard matrix-vector multiplication, which scales quadratically with the matrix size. In
practice, highly optimized linear tensor algebra libraries can be used to perform the tensor contraction such as the \href{https://julialang.org/}{\texttt{Julia}} package \href{https://jutho.github.io/TensorOperations.jl/}{\texttt{TensorOperations}} \cite{jutho2019}.

\subsection{Kronecker sums} Let $\mmat{A}_k, \; k=1, \ldots , d$, denote a set of $d$ square matrices. The Kronecker sum is generally defined for two square matrices as $\mmat{A}_1 \oplus \mmat{A}_2 := \mmat{I}_2 \otimes \mmat{A}_1 + \mmat{A}_2 \otimes \mmat{I}_1$. More generally, we may define
\begin{align*}
	\mmat{A}_1 \oplus \mmat{A}_2 \oplus \ldots \oplus \mmat{A}_d :&= \mmat{I}_d \otimes \ldots \otimes \mmat{I}_2 \otimes \mmat{A}_1 	\\
				&+  \mmat{I}_d \otimes  \ldots \otimes \mmat{A}_2 \otimes \mmat{I}_1 	\\ 
				&+ \vdots 																							\\
				&+  \mmat{A}_d \otimes \ldots \otimes \mmat{I}_2 \otimes \mmat{I}_1.
\end{align*}
The Kronecker sum is invariant under permutations. However, it is not associative and also not distributive with respect to the Kronecker product. 

Let $\mmat{M}_k, \; k=1, \ldots , d$, denote a set of $d$ symmetric positive definite matrices, e.g. mass matrices. We define the following generalized Kronecker sum,
\begin{align*}
	\mmat{A}_1 \oplushat \mmat{A}_2 \oplushat   \ldots   \oplushat \mmat{A}_d :&= \mmat{M}_d \otimes \ldots \otimes \mmat{M}_2 \otimes \mmat{A}_1 	\\
				&+  \mmat{M}_d \otimes  \ldots \otimes \mmat{A}_2 \otimes \mmat{M}_1 	\\ 
				&+ \vdots 																							\\
				&+  \mmat{A}_d \otimes \ldots \otimes \mmat{M}_2 \otimes \mmat{M}_1.
\end{align*}
A generalized Kronecker sum can always be transformed into a regular Kronecker sum. Consider the Cholesky factorization of the matrix $\mat{M} = \mmat{M}_d \otimes \ldots \otimes \mmat{M}_1$, 
\begin{align*}
	\mat{M} = \mat{L} \mat{L}^T = \left(\mmat{L}_d \otimes \ldots \otimes \mmat{L}_1 \right) \left(\mmat{L}^T_d \otimes \ldots \otimes \mmat{L}^T_1 \right).
\end{align*}
Using the mixed product property of Kronecker product matrices, we may write
\begin{align*}
	\mat{L}^{-1} \left(\mmat{A}_1 \oplushat \mmat{A}_2 \oplushat   \ldots   \oplushat \mmat{A}_d \right) \mat{L}^{-T} =
	\tilde{\mmat{A}}_1 \oplus \tilde{\mmat{A}}_2 \oplus   \ldots   \oplus \tilde{\mmat{A}}_d,
\end{align*}
where $ \tilde{\mmat{A}}_k = \mmat{L}_k^{-1} \mmat{A}_k  \mmat{L}_k^{-T}$, $k=1, \ldots , d$.

\subsection{Fast diagonalization of Kronecker sums}

Let $\mmat{A}_k = \mmat{U}_{k} \mmat{\Lambda}_k  \mmat{U}^{-1}_{k}$ denote the eigenvalue decomposition of matrices $\mmat{A}_k, \; k=1,\ldots , d$. Using $\mmat{I}_k = \mmat{U}_{k} \mmat{U}^{-1}_{k}$ and repeated application of the mixed-product property of Kronecker product matrices, it is easy to show that a Kronecker sum has the following eigenvalue decomposition,
\begin{align}
	\mmat{A}_1 \oplus \mmat{A}_2 \oplus \ldots \oplus \mmat{A}_d = \left(\mmat{U}_d \otimes \ldots \otimes \mmat{U}_1 \right) 
					\left(\mmat{\Lambda}_1 \oplus \ldots \oplus \mmat{\Lambda}_d \right) 
					 \left(\mmat{U}^{-1}_d \otimes \ldots \otimes \mmat{U}^{-1}_1 \right).
\end{align}
Similarly, the eigenvalue decomposition of a generalized Kronecker sum is
\begin{align*}
	\mmat{A}_1 \oplushat \mmat{A}_2 \oplushat   \ldots   \oplushat \mmat{A}_d = \left(\tilde{\mmat{U}}_d \otimes \ldots \otimes \tilde{\mmat{U}}_1 \right) 
					\left(\mmat{\Lambda}_1 \oplus \ldots \oplus \mmat{\Lambda}_d \right) 
					 \left(\tilde{\mmat{U}}^{-1}_d \otimes \ldots \otimes \tilde{\mmat{U}}^{-1}_1 \right) 
\end{align*}
with $\tilde{\mmat{U}}_k = \mmat{L}^{-T}_k \mmat{U}$, $k=1,\ldots,d$.

Since the decompositions involve Kronecker product matrices and diagonal matrices the inverse matrix can be constructed efficiently using the Kronecker sum eigendecomposition.

%% file: sections/appendix2_discretization.tex
\section{Structure preserving isogeometric discretization}
\label{app:discretization}

\subsection{Univariate splines} A spline is a piecewise polynomial that is characterized by the polynomial degree of its segments and the regularity prescribed at their interfaces. Consider a partitioning $\Delta$ of the univariate interval $[a,b]$ into a sequence of breakpoints,
\begin{align}
	a = \hat{x}_0 < \hat{x}_1 < \ldots < \hat{x}_m = b.
\end{align}
With every internal breakpoint, $\hat{x}_k$, we may associate an integer, $r_k$, prescribing the smoothness between the polynomial pieces. 

Using e.g. the Cox–DeBoor recursion \cite{CoHuBa2009,PiTi1995} it is possible to construct $n$ linearly independent B-spline basis functions, $\List{N_{i,p}(\hat{x}), \; i=1, \ldots, n}$. These functions span the space of smooth splines of polynomial degree $p$ and smoothness $r_k$ at breakpoint $\hat{x}_k$ defined over $\Delta$,
\begin{align}
	\mathbb{S}^{\p}_\r (\Delta) := \Span{N_{i,p}(\hat{x}), \; i=1, \ldots, n}.
\end{align}
B-splines have important mathematical properties, many of which are useful in design as well as in analysis. B-spline basis functions of degree $p$ may have up to $p-1$ continuous derivatives, they form a positive partition of unity, and have local support of up to $p+1$ elements. 

An important property with respect to differentiation is the following,
\begin{align}
	\frac{d}{d\hat{x}} \sum_{i=1}^n \alpha_i \; N_{i,p}(\hat{x}) = \sum_{i=1}^{n-1} \gamma_i \; M_{i,p-1}(\hat{x}), \quad
	\gamma_i = \alpha_{i+1} - \alpha_{i}.
	\label{eq:reduction}
\end{align}
The functions $\List{M_{i,p-1}(\hat{x}), \; i=1, \ldots, n-1}$ denote a special normalization of B-splines that have unit integral, that is,
\begin{align}
	M_{i,p-1}(\hat{x}) = \frac{N_{i,p-1}(\hat{x})}{\int N_{i,p-1}}.
\end{align}
Differentiation of splines, $\frac{d}{dx} \sspace{\p}{\r} \mapsto \sspace{\p-1}{\r-1}$, may now be encoded by discrete differentiation of the coefficients, that is,
\begin{align}
	\begin{bmatrix}
		\gamma_1 \\
		\gamma_2 \\
		\vdots 			\\
		\gamma_{n-1}
	\end{bmatrix} = \underbrace{
	\begin{bmatrix}
		-1 	& 1			&				&			&		\\
		  		& -1			&	1			&			&		\\
		  		&	\ddots	&	\ddots	&			&		\\
		  		& 				&				&	-1 	& 1
	\end{bmatrix}	}_{\mmat{D}_{n}}
	\begin{bmatrix}
		\alpha_1 			\\
		\alpha_2 			\\
		\vdots 				\\
		\alpha_{n-1} 	\\
		\alpha_n
	\end{bmatrix}.
\end{align}

\subsection{A discrete Stokes complex}
Splines have proven to be highly effective as trial and test spaces in many applications. Multidimensional spaces are most conveniently constructed using the tensor product. For example, in $\mathbb{R}^3$ we may define on $\Omega := \domain_1 \times \domain_2 \times \domain_3 \subset \mathbb{R}^3$ the tensor product spline space
\begin{align}
	\sspace{\p_1,\p_2,\p_3}{\r_1,\r_2,\r_3}(\Omega)  := \sspace{\p_1}{\r_1}(\domain_1) \otimes \sspace{\p_2}{\r_2}(\domain_2) \otimes \sspace{\p_3}{\r_3}(\domain_3).
\end{align}
Using tensor product splines it is possible to construct discrete velocity and pressure spaces that are in some way naturally compatible with one another
\begin{align}
	V_h 		&:= \sspace{\p_1,\p_2-1,\p_3-1}{\r_1,\r_2-1,\r_3-1}(\Omega) \times
        			\sspace{\p_1-1,\p_2,\p_3-1}{\r_1-1,\r_2,\r_3-1}(\Omega) \times
        			\sspace{\p_1-1,\p_2-1,\p_3}{\r_1-1,\r_2-1,\r_3}(\Omega),					\\
	Q_h 		&:= \sspace{\p_1-1,\p_2-1,\p_3-1}{\r_1-1,\r_2-1,\r_3-1}(\Omega).
\end{align}
Let $V := \vect{H}^1(\Omega)$ and $Q := L^2(\Omega)$. It follows that $V_h \subset V$ and $Q_h \subset Q$ form a discrete Stokes complex via the commuting diagram
\begin{equation}
\begin{tikzcd}
\arrow[d, "\Pi^h_V"]
  V \arrow[r, "\nabla \cdot"] & \arrow[d, "\Pi^h_Q"] Q \\
  V_h \arrow[r, "\nabla \cdot"] & Q_h
\end{tikzcd}
\end{equation}
Here, $\Pi^h_V$ and $\Pi^h_Q$ are suitable projection operators that can be used to proof inf-sup stability and consistency, see  \cite{BuFaSa2010, EvHu2013}. The spaces naturally satisfy a discrete divergence theorem, which leads to point-wise divergence-free velocity fields in discretizations of incompressible Stokes and Navier-Stokes flow \cite{BuFaSa2010,EvHu2013, HiToHuGe2014}.

\subsection{Definitions of univariate matrices on Cartesian grids}
\label{subsec:univariatemats}
In the following we use the notation $N_{i}(\hat{x}) := N_{i,p}(\hat{x})$ and $M_{i}(\hat{x}) := M_{i,p-1}(\hat{x})$ and write $\phi^{\prime}(\hat{x}): = \tfrac{d}{d\hat{x}} \phi (\hat{x})$ to denote differentiation. This allows a concise definition of the integrals involved in our computations,
\begin{align*}
	&        \mmat{M}_{ij} = \int_0^1 N_{i}(\hat x) N_{j}(\hat x) \,\mathrm d \hat x, 
   &&        \mmat{K}_{ij} = \int_0^1 N^\prime_{i}(\hat x) N^\prime_{j}(\hat x) \,\mathrm d \hat x, 
   &&        \mmat{C}_{ij} = \int_0^1 N_{i}(\hat x) M^\prime_{j}(\hat x) \,\mathrm d \hat x, \\
	&\mmat{\check{M}}_{ij} = \int_0^1 M_{i}(\hat x) M_{j}(\hat x) \,\mathrm d \hat x, 
   &&\mmat{\check{K}}_{ij} = \int_0^1 M^\prime_{i}(\hat x) M^\prime_{j}(\hat x) \,\mathrm d \hat x, 
   &&\mmat{\check{C}}_{ij} = \int_0^1 N^\prime_{i}(\hat x) M_{j}(\hat x) \,\mathrm d \hat x. 
\end{align*}
For the contributions of the boundary terms, we additionally define
\begin{align*}
	&[\mmat{\check{N}}]_{ij} = M_{i}(1) M_{j}(1) + M_{i}(0) M_{j}(0)
	&& \text{and}
	&&[\mmat{\check{B}}]_{ij} = M_{i}(1) M^\prime_{j}(1) - M_{i}(0) M^\prime_{j}(0).
\end{align*}

%% file: ex_article.bbl
\begin{thebibliography}{10}

\bibitem{antolin2015efficient}
{\sc P.~Antolin, A.~Buffa, F.~Calabro, M.~Martinelli, and G.~Sangalli}, {\em Efficient matrix computation for tensor-product isogeometric analysis: The use of sum factorization}, Computer Methods in Applied Mechanics and Engineering, 285 (2015), pp.~817--828.

\bibitem{arnold2018finite}
{\sc D.~N. Arnold}, {\em Finite element exterior calculus}, SIAM, 2018.

\bibitem{BeIsAd1965}
{\sc A.~Ben-Israel}, {\em An iterative method for computing the generalized inverse of an arbitrary matrix}, Mathematics of Computation - Math. Comput., 19 (1965), pp.~452--452.

\bibitem{ben2003generalized}
{\sc A.~Ben-Israel and T.~N. Greville}, {\em Generalized inverses: theory and applications}, vol.~15, Springer Science \& Business Media, 2003.

\bibitem{benzi2002preconditioning}
{\sc M.~Benzi}, {\em Preconditioning techniques for large linear systems: a survey}, Journal of computational Physics, 182 (2002), pp.~418--477.

\bibitem{benzi2008saddlepoint}
{\sc M.~Benzi and A.~J. Wathen}, {\em Some preconditioning techniques for saddle point problems}, Model order reduction: theory, research aspects and applications,  (2008), pp.~195--211.

\bibitem{BrTa2018}
{\sc A.~Bressan and S.~Takacs}, {\em Sum-factorization techniques in isogeometric analysis}.
\newblock 2018, \url{https://hal.science/hal-01874006}.

\bibitem{BuFaSa2010}
{\sc A.~Buffa, C.~de~Falco, and G.~Sangalli}, {\em Isogeometric analysis: Stable elements for the 2d stokes equation}, International Journal for Numerical Methods in Fluids, 65 (2011), pp.~1407--1422.

\bibitem{calabro2017fast}
{\sc F.~Calabro, G.~Sangalli, and M.~Tani}, {\em Fast formation of isogeometric galerkin matrices by weighted quadrature}, Computer Methods in Applied Mechanics and Engineering, 316 (2017), pp.~606--622.

\bibitem{chen2005matrix}
{\sc K.~Chen}, {\em Matrix preconditioning techniques and applications}, vol.~19, Cambridge University Press, 2005.

\bibitem{ClThWe2001}
{\sc J.-J. Climent, N.~Thome, and Y.~Wei}, {\em A geometrical approach on generalized inverses by neumann-type series}, Linear Algebra and its Applications, 332-334 (2001), pp.~533--540.

\bibitem{CoHuBa2009}
{\sc J.~Cottrell, T.~Hughes, and Y.~Bazilevs}, {\em Isogeometric Analysis: Toward integration of CAD and FEA}, 09 2009.

\bibitem{DuGrRo1979}
{\sc P.~F. Dubois, A.~Greenbaum, and G.~H. Rodrigue}, {\em Approximating the inverse of a matrix for use in iterative algorithms on vector processors}, Computing, 22 (1979), pp.~257--268.

\bibitem{EvHu2013}
{\sc J.~A. Evans and T.~J.~R. Hughes}, {\em Isogeometric divergence-conforming b-splines for the darcy-stokes-brinkman equations}, Mathematical Models and Methods in Applied Sciences, 23 (2013), pp.~671--741.

\bibitem{HiToHuGe2014}
{\sc R.~Hiemstra, D.~Toshniwal, R.~Huijsmans, and M.~Gerritsma}, {\em High order geometric methods with exact conservation properties}, Journal of Computational Physics, 257 (2014), pp.~1444--1471.

\bibitem{HiSaTaCaHu2019}
{\sc R.~R. Hiemstra, G.~Sangalli, M.~Tani, F.~Calabrò, and T.~J. Hughes}, {\em Fast formation and assembly of finite element matrices with application to isogeometric linear elasticity}, Computer Methods in Applied Mechanics and Engineering, 355 (2019), pp.~234--260.

\bibitem{householder2013theory}
{\sc A.~S. Householder}, {\em The theory of matrices in numerical analysis}, Courier Corporation, 2013.

\bibitem{JoMiChaGe1983}
{\sc O.~G. Johnson, C.~A. Micchelli, and G.~Paul}, {\em Polynomial preconditioners for conjugate gradient calculations}, SIAM Journal on Numerical Analysis, 20 (1983), pp.~362--376.

\bibitem{jutho2019}
{\sc Jutho, getzdan, S.~Lyon, M.~Protter, M.~P. S, Leo, J.~Garrison, F.~Otto, E.~Saba, D.~Iouchtchenko, A.~Privett, and A.~Morley}, {\em Jutho/tensoroperations.jl: v1.1.0}, June 2019.

\bibitem{Loan2000}
{\sc C.~F. Loan}, {\em The ubiquitous {K}ronecker product}, Journal of Computational and Applied Mathematics, 123 (2000), pp.~85--100.

\bibitem{MoSaTa2018}
{\sc M.~Montardini, G.~Sangalli, and M.~Tani}, {\em Robust isogeometric preconditioners for the stokes system based on the fast diagonalization method}, Computer Methods in Applied Mechanics and Engineering, 338 (2018), pp.~162--185.

\bibitem{OLeary1991}
{\sc D.~P. O'Leary}, {\em Yet another polynomial preconditioner for the conjugate gradient algorithm}, Linear Algebra and its Applications, 154-156 (1991), pp.~377--388.

\bibitem{PeSt2011}
{\sc M.~Petkovic and P.~Stanimirovic}, {\em Iterative method for computing the {M}oore–{P}enrose inverse based on {P}enrose equations}, Journal of Comp. and Applied Mathematics, 235 (2011).

\bibitem{petkovic2015hyper}
{\sc M.~D. Petkovi{\'c} and M.~S. Petkovi{\'c}}, {\em Hyper-power methods for the computation of outer inverses}, Journal of computational and applied mathematics, 278 (2015), pp.~110--118.

\bibitem{PiTi1995}
{\sc L.~A. Piegl and W.~Tiller}, {\em The \uppercase{NURBS} book}, in Monographs in Visual Communications, 1995.

\bibitem{Saad1985}
{\sc Y.~Saad}, {\em Practical use of polynomial preconditionings for the conjugate gradient method}, SIAM Journal on Scientific and Statistical Computing, 6 (1985), pp.~865--881.

\bibitem{Saad1989}
{\sc Y.~Saad}, {\em Krylov subspace methods on supercomputers}, SIAM Journal on Scientific and Statistical Computing, 10 (1989), pp.~1200--1232.

\bibitem{saad2003iterative}
{\sc Y.~Saad}, {\em Iterative methods for sparse linear systems}, SIAM, 2003.

\bibitem{Schulz1933}
{\sc G.~Schulz}, {\em Iterative berechung der reziproken matrix}, Zeitschrift für Angewandte Mathematik und Mechanik, 13 (1933), pp.~57--59.

\bibitem{Tanabe1975}
{\sc K.~Tanabe}, {\em Neumann-type expansion of reflexive generalized inverses of a matrix and the hyperpower iterative method}, Linear Algebra and its Applications, 10 (1975), pp.~163--175.

\bibitem{vorst2003iterative}
{\sc H.~A. Van~der Vorst}, {\em Iterative Krylov methods for large linear systems}, no.~13, Cambridge University Press, 2003.

\bibitem{wathen2015preconditioning}
{\sc A.~J. Wathen}, {\em Preconditioning}, Acta Numerica, 24 (2015), pp.~329--376.

\end{thebibliography}
